**A low-tech solution to process entire metal/molecule heterostructure stacks into vertical nanopillar electronic devices.**

T. Zafar[1]*, L. M. Kandpal[1]*, E. Urbain[1]*, N. Beyer[1], B. Gobaut[1], L. Joly[1], H. Majjad[1], S. Siegwald[1], D. Mertz[1], B. Leconte[1], C. Kieber[1], V. Da Costa[1], W. Weber[1], S. Boukari[1], M. Bowen[1]

*: equal contributions

[1]Institut de Physique et Chimie des Matériaux de Strasbourg (IPCMS), UMR 7504 CNRS, Université de Strasbourg, 23 Rue du Lœss, BP 43, 67034, Strasbourg, France.

*Corresponding author:

Institut de Physique et Chimie des Matériaux de Strasbourg (IPCMS), UMR 7504 CNRS, Université de Strasbourg, 23 Rue du Lœss, BP 43, 67034 Strasbourg, France. (email: bowen@unistra.fr; talhazafarali@gmail.com; lmkandpal09@gmail.com)

# Contents

**ABSTRACT:**

Quantum technologies aim to assemble devices whose operation is controlled by the quantum state of individual atoms. Achieving this level of control in a practical, scalable design remains, however, a major obstacle to mass societal adoption. By working at the level of interatomic bonding, molecular engineering has enabled exquisite control over the electronic properties of individual atoms and their interactions with neighboring atoms. This positions molecular electronics as a potentially disruptive quantum technology, but serious technological challenges have prevented it from being included in technical road maps. The main obstacle is that conventional, mass scalable nanodevice technologies utilize resists and solvents that can degrade molecules. Some approaches involve exposing junction interfaces to contaminants (e.g. air, resist etc…), which can be particularly problematic for spintronics. In this technical paper, we present our decade-long work into building a nanotechnological chain that can process entire metal/molecule heterostructures into vertical nanopillars electronic devices. We discuss the advantages and pitfalls of the various iterations of this process that were implemented. We also discuss outlooks for this unique technology.

Keywords: Nanolithography; Nanoscale patterning; Organic thin films; Micro- and nano devices; Quantum effects; Organic magnetic tunnel junctions; Spintronic; Spinterface.

# 1. Intro

Creating sophisticated functional materials is made possible by the capacity to design molecular structures at the atomic level. By precisely controlling atoms and their environment, molecular engineering makes it possible to create nanostructures with specific optical, magnetic, and electrical behaviors. Magnetic tunnel junctions (MTJs)[1] have gained wide popularity due to their mature commercialization in tunnel magnetoresistance (TMR) read-heads[2] within the computer hard-disks and non-volatile random access memories[3] (RAM) based on the spin-transfer torque (STT) effect[1,4–7] . Beyond conventional storage devices, MTJs have enabled the experimental realization of STT-based high speed and power efficient embedded static random-access memory (S-RAM) for logics (microprocessors)[8–10], emerging non-volatile hybrid storage class memory (SCM) with comparable speed of standalone dynamic random-access memory (DRAM)[11], a non-von-Neumann architecture based artificial neurons[12–14], radiation immune electronics[15,16], and radio-frequency (rf) emitter and detectors[4,5,17–20].

The conventional MTJ is comprised of an ultrathin MgO layer serving as the tunnel barrier. Efforts to replace it with a thicker organic or molecular spacer layer were based on the premise that diffusive transport across lighter atoms would produce a weak spin-orbit coupling ($\propto Z^4$). The weak spin-orbit coupling (SOC), together with the electronic transport through delocalized $\pi$-molecular orbitals, would offer large spin lifetimes ($\tau s$)[21] for the injected spin-polarized electrons. Note here that the spin lifetime ($\tau s$) for these organic molecules can reach up to the milliseconds[22–24]. The absence of a nuclear magnetic moment in the carbon atom further rules out the possibility of a hyperfine interaction in these molecules thereby preserving spin coherence of injected spin-polarized electrons over comparatively large transport distances. Furthermore, in the tunnel transport regime, organic MTJs (OMTJs)[25] enable the exploration of quantum phenomena, such as electron transport through spin chains[26]. Beyond their superior spin-preserving characteristics, organic molecules offer remarkable tenability by allowing their structural, electronic, and chemical properties to be precisely engineered through chemistry. This can in turn reduce the production cost and scalability of OMTJs[27,28]. These aforementioned advantages of OMTJs position them as a potential candidate for next generation spin-electronic[29] (spintronic) devices.

The fabrication of high-quality OMTJ nanopillars is challenging due to the complex techniques and specialized apparatus required. The most frequent technique to produce MTJ nanopillars is electron beam lithography (EBL) using the top-down[30,31] or stencil approaches[32], while other techniques such as deep ultraviolet photolithography (DUV)[33] and focused-ion beam (FIB) have also been used[34]. Nonetheless, these lithographic approaches involve chemical treatment during the development and liftoff stages, which can destroy the molecular spacer.

Currently there are two main techniques used to fabricate organic magnetic tunnel junctions. 1) The "shadowmask junction'' technique relies on the use of shadow masks. In this method, both electrodes and the barrier layer are deposited sequentially through a series of shadow masks, forming a cross pattern. The tunnel junction is created at the intersection of the cross. Advantageously, all depositions can be performed under vacuum, thereby promoting high quality interfaces. However, due issues of mask structural stability, the mask's effective thickness and possible shadowing effects, this technique yields mostly macroscale junctions (typically 100x100 $\mu m^2$)[35–37], though the IPR Rennes group has managed to achieve 5x5 $\mu m^2$ junctions[37]. 2) The second technique, implemented by the LAF Paris and IJL Nancy groups[38,39], utilizes a technological layer to circumscribe the junction surface area upon later growth of the counterelectrode, and potentially also of the organic layer. After depositing the bottom electrode, the sample is removed to air and a thin dielectric (resist, SiOx…) is deposited onto the entire sample. Openings in this technological layer are then performed (e.g. using UV lithography or with the tip of an atomic force microscope[38] (AFM) ). The hole is then filled with the top electrode. It is a dry process technology that enables the production of nano-scale vertical junctions without the use of solvents. However, the junction technological steps, which are undertaken between the deposition of the bottom and top MTJ electrodes, can cause potential degradation to the final junction: the exposure of the bottom interface to air must be addressed[38,40,41], and structural damage to the molecular layer upon nanoindentation can occur.

What is missing from this landscape is a process that takes a full OMTJ stack, grown without breaking vacuum, and transform it into sub-micronic vertical pillars. We've addressed this challenge by developing a process based on the masking properties of micro- or nanometric beads as a shadow mask for the pillar itself, while the lower and top electrodes are shaped using shadow masks. Early on in the process development, the silica nanobead positioning was accomplished using a dry approach that used ultra-short ultrasonic power pulses to disperse beads onto the stack. To control the number of nanobeads and their location within a macrojunction, we turned to a microdroplet printing technique[25,42]. In this technical paper, we begin with a brief overview of the components of MTJs and the construction of OMTJs, followed by a detailed description of a new wet technique and its possible variations. This work aims to provide an accessible and scalable solution for researchers and industries seeking to explore advanced organic electronic devices at reduced costs and with greater ease.

## 2. Process Overview

To guide the reader through the contents of this technical paper, we begin with this overview of the technological process.

### 2.1. Step 1: Heterostructure deposition into lower electrode form

The first process step is to deposit the entire heterostructure through a hard shadow mask into lower electrode patterns. This deposition is performed in a UHV deposition cluster. A schematic of the lower-electrode shaped deposition is shown in

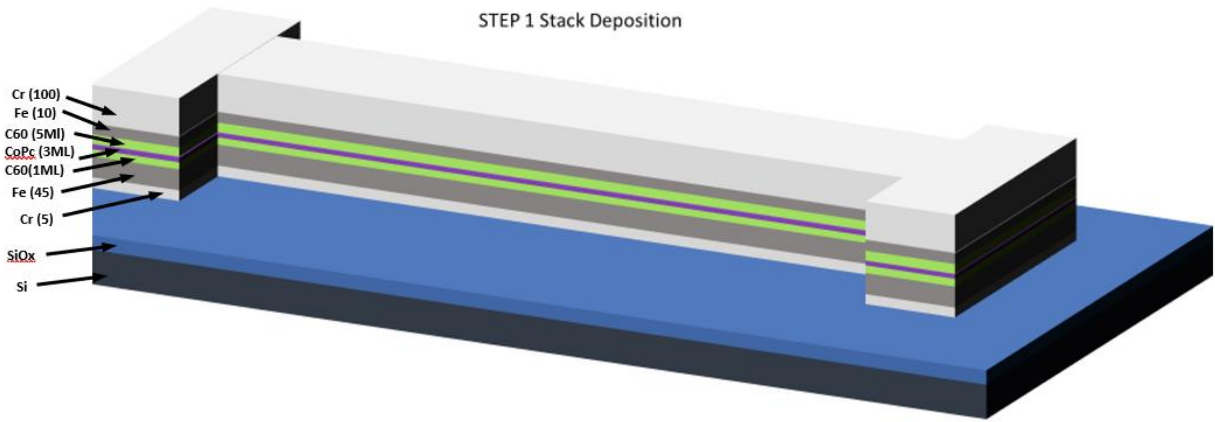

Figure 1

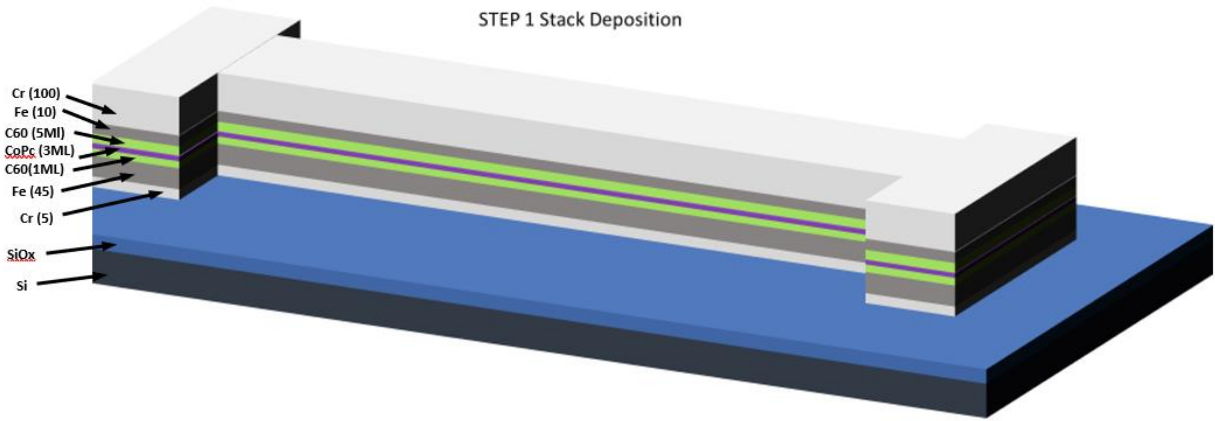

*Figure 1: Step 1: Thin film heterostructure stack deposited through Mask 1, into the shape of the lower electrode.*

## 2.2. Step 2: Nanopillar masking using nanobeads

The technically tricky part of the process is now to obtain a mask against which to etch the heterostructure into nanopillars form, without solvent/developer baths. To accomplish this key process step, we deposit silica nanobeads on the heterostructure. The *idealized* result is schematized in Figure 2.

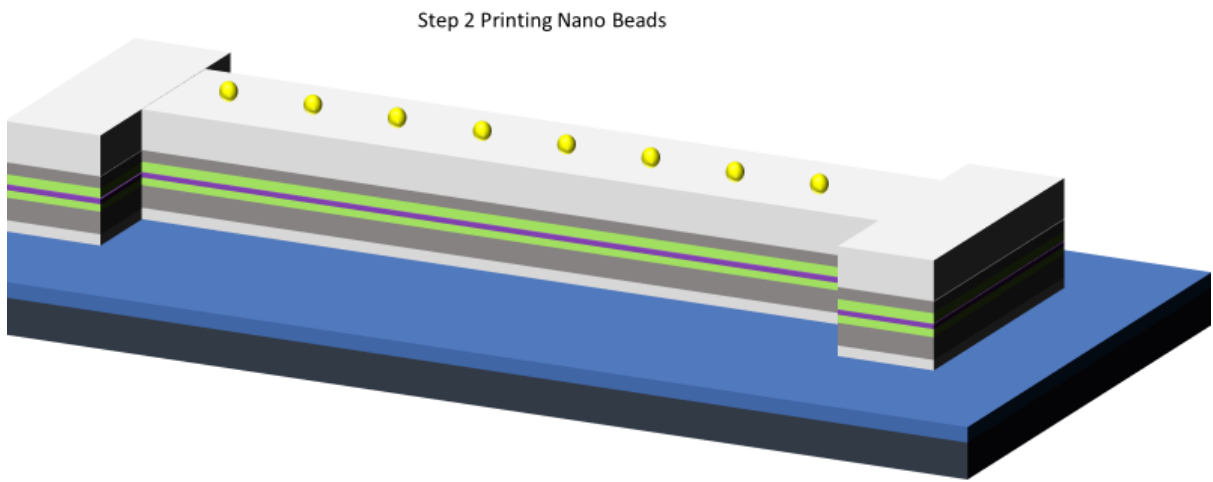

*Figure 2: Step 2: Silica nanobeads are deposited onto the heterostructure of Step 1, to serve as a mask against etching (Step 3), in order to define nanopillars.*

## 2.3. Step 3: Nanopillar Etching and Dielectric Encapsulation

In Step 3, the sample is etched down to the lower electrode (see Figure 3), and is encapsulated with a dielectric (see Figure 4). Note how the end areas of the lower electrodes are protected from etching and encapsulation, using a hard mask, in order to enable an electrical contact onto the lower electrode across the stack.

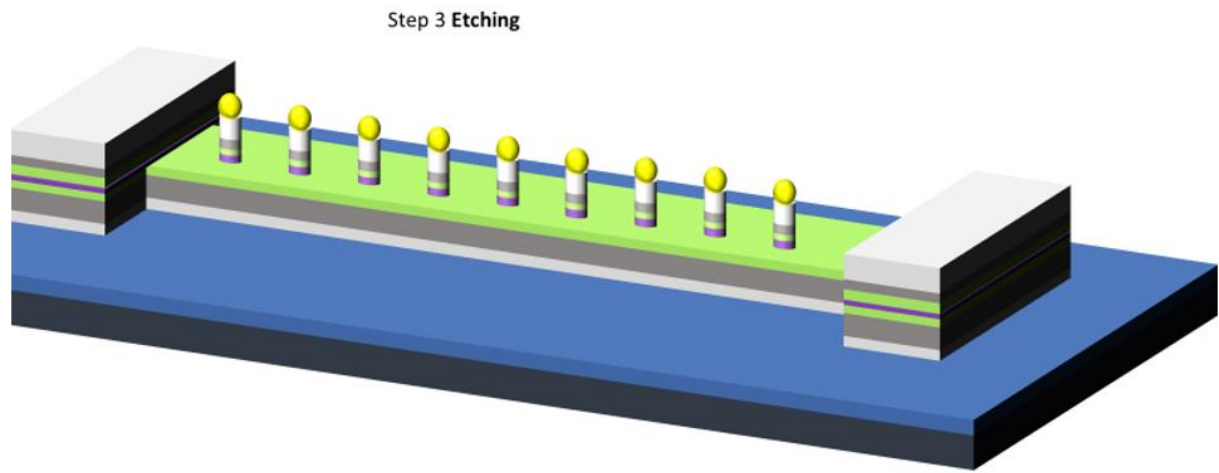

*Figure 3: Step 3a: Heterostructure etching down to the lower electrode. Top electrode material is retained under the nanobeads, and on the end areas of the lower electrode thanks to hard mask 2.*

.

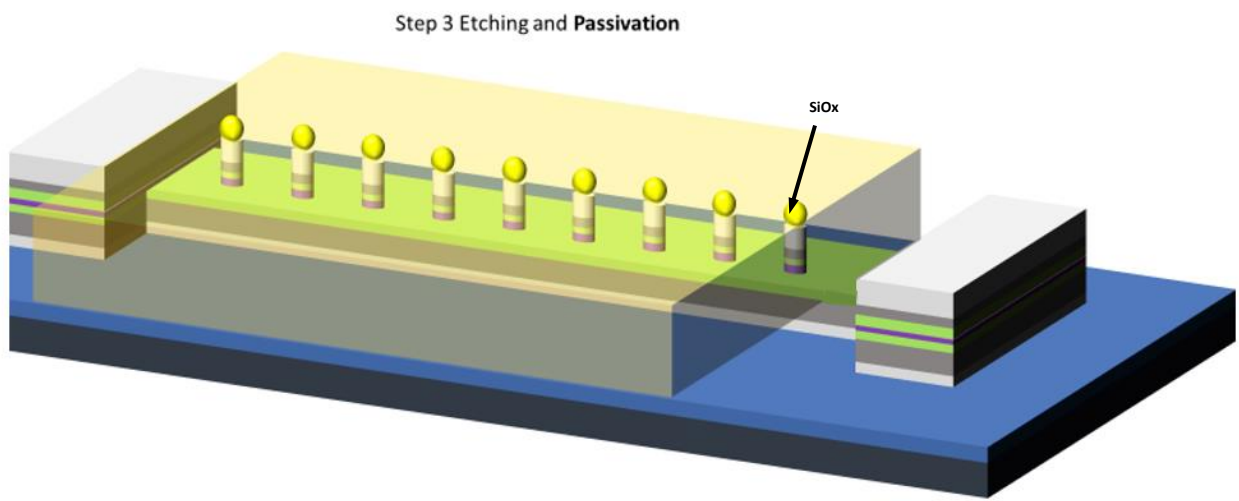

*Figure 4: Step 3b: Dielectric encapsulation of the sample while the hard mask 2 is still in place. Dielectric $SiO_2$ is deposited everywhere except on the side pads.*

## 2.4. Step 4: Nanobead Liftoff

To remove the nanobeads from the pillar tops as shown in Figure 5, some form of mechanical energy is applied (see section 7.2 for details).

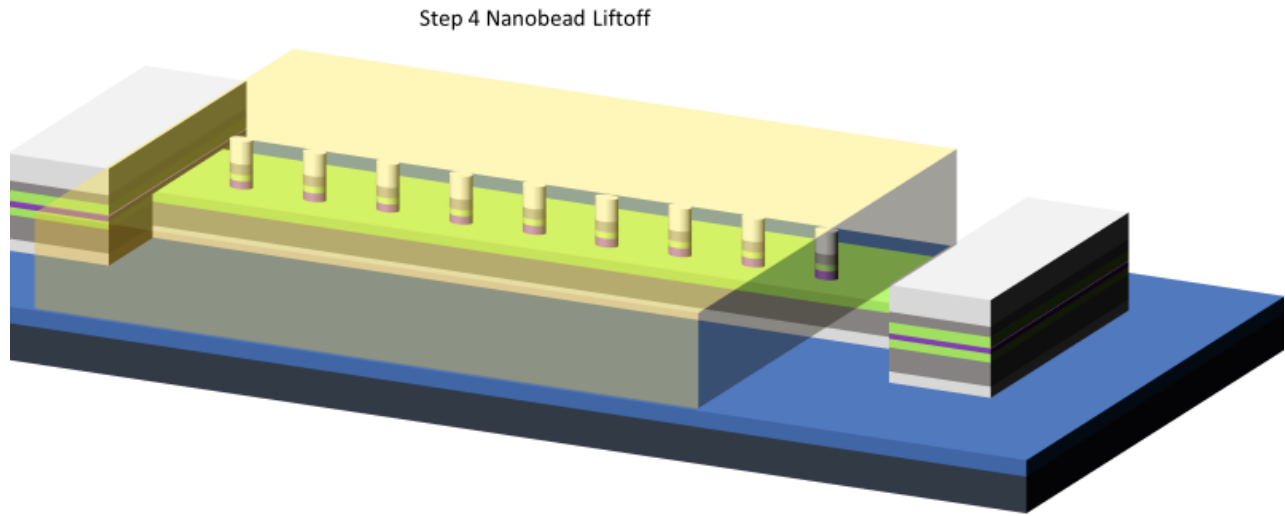

*Figure 5: Step 4: the nanobeads are removed from the sample surface using some form of mechanical energy See Section 7.2 for details.*

## 2.5. Step 5: Top Electrode Metallization

In the final Step 5 of the process, top electrode contacts and bottom electrode bonding pads are defined by depositing metals through a final hard mask 3 (see Figure 6).

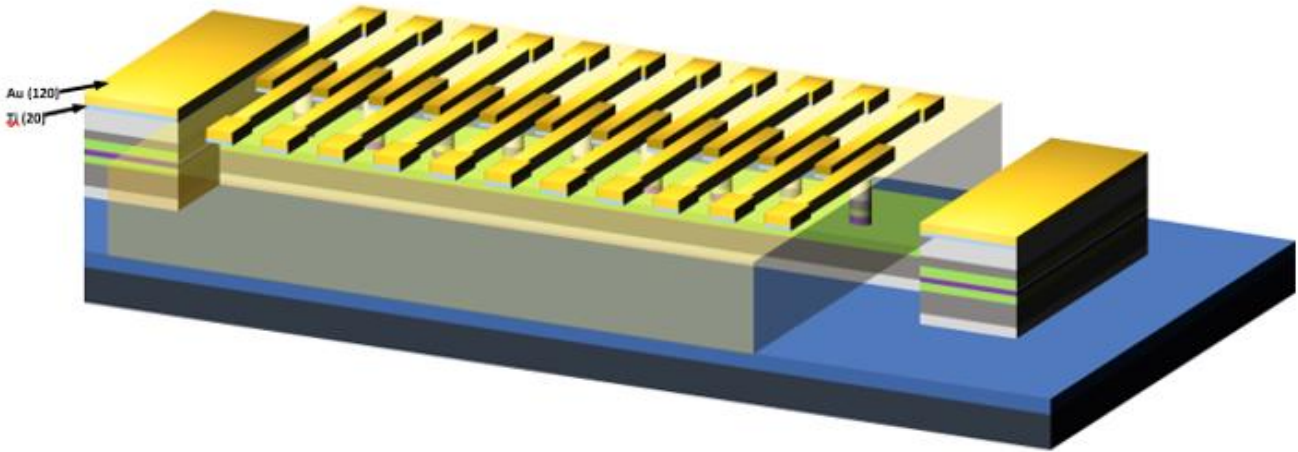

Figure 6: Step 5: Definition of top/bottom electrode contacts by depositing metals through a hard mask 3.

## 2.6. Thickness Considerations

As with all vertical junction processes, there exists a relationship between the thicknesses of the heterostructure's top and bottom metallic layers $dLE$ and $dTE$, of the dielectric and metallization layers $dD$ and $dM$, and the radius R of the nanobeads:

$$dM > R > dD > dLE, dTE$$

In order for the nano bead lift off to work (Step 4), the dielectric thickness $dD$ must be less than the radius R of the silica nanobeads. The top contact of the nanopillar must be made into the hole in the dielectric created by the nanobead removal, so $dM > dD$. Finally, the dielectric should at least cover the lower electrode, otherwise a short-circuit can occur, so $dD > dLE$ (note: a factor of 1.5 would constitute a better guarantee). The condition $dD > dTE$ ensures that the pillar is entirely passivated, and its necessity depends on the impact of exposing these materials to ambient conditions. These considerations are schematized in Figure 7.

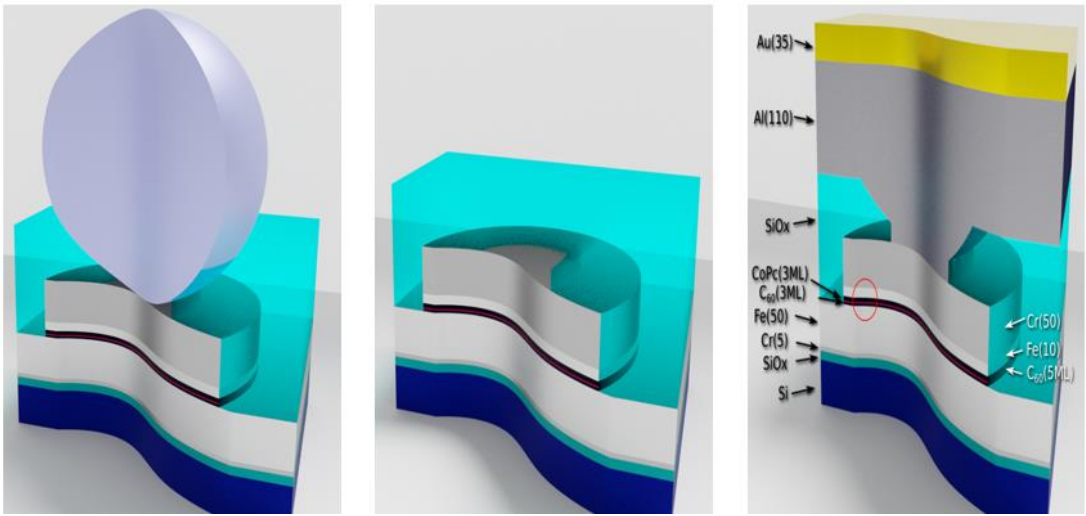

*Figure 7: Schematic illustration of the fabrication process for vertical organic MTJs using silica nanobeads. Steps include (left) heterostructure deposition nanobead printing, etching, and dielectric deposition with thickness smaller than the bead radius to enable (middle) the subsequent bead removal. (right) The top contact fills the bead void, while the dielectric isolates the bottom electrode to prevent short-circuits.*

# 3. Heterostructure deposition

In this Section, we shall first describe details of our heterostructure growth, using the example of stacks used for energy harvesting[25], with a focus on the nanojunction technology that is built into what is otherwise a materials science step.

## 3.1. Materials Science

The Si/SiOx//Cr(5)/Fe(45)/$C_{60}$(2ML)/CoPc(3ML)/$C_{60}$(5ML)/Fe(15)/Cr(100) stack deposition (numbers in nm unless otherwise noted) is done in a ultra-high vacuum (UHV) multi-chamber cluster. Note here the very thick Cr capping layer, which was found to be crucial in order to avoid oxygen interdiffusion to the lower electrode[43]. A robotic arm in the transfer chamber with base pressure $3 \times 10^{-10}$ mbar enables the transfer between sputter, metal sublimation and molecular (OMBE) sublimation chambers. Before the thin film heterostructure is deposited, the substrate is annealed to 160°C for 2 hours and then cooled to room temperature naturally in 5 hours. Fe is used as the ferromagnetic (FM) electrodes, and Cr is used as a substrate buffer layer and as a capping layer. These metals are sputtered at ambient temperature with an Ar pressure of $3 \times 10^{-4}$ mbar. The base pressure within the chamber is $1 \times 10^{-9}$ mbar. 10 minutes are needed to transfer from the sputtering chamber to the OMBE chamber (base pressure of $5 \times 10^{-10}$ mbar). The CoPc and $C_{60}$ films were grown at room temperature at a pressure of $2 \times 10^{-9}$ mbar. Lastly, the samples are returned via the robotic arm to the sputtering chamber to deposit the Fe/Cr counter electrode. Table 1 provides a detailed information of the deposited layers and the parameters that were used.

| Material | Function in Stack | Deposition Method | Deposition Pressure (mbar) | Rate of deposition |
|---|---|---|---|---|
| Fe | Ferromagnetic Electrode | Magnetron Sputtering | $5.4 \times 10^{-4}$ | 0.2 A°/s |
| Cr | Buffer and Capping | Magnetron Sputtering | $7.33 \times 10^{-4}$ | 0.3 A°/s |
| $C_{60}$ | Insulating amorphous molecular carrier | Molecular epitaxy | $3 \times 10^{-9}$ | 0.006 ML/s |
| CoPc | Insulating molecular barrier and source of PM center | Molecular epitaxy | $2 \times 10^{-9}$ | 0.008 ML/s |

*Table 1: Deposition processes and parameters for the various materials used to create organic hybrid heterostructures. All depositions were done in situ with a sample at nominally room temperature.*

## 3.2. Nanojunction Technology

### 3.2.1. Loose Mask 1 Implementation

In a first iteration of this process, we simply affixed Al sheets with a cutout of the shadow-mask growth required to make lower and upper electrodes. The requirement to hand align each mask and maintain it onto the sample using pins is possible, as shown in Figure 8. An optical microscopic image of the resulting sample growth is shown in Figure 9 . This technique can lead to misalignment's: extreme examples are shown in Figure 100. This approach is not only time-consuming, but also prone to errors, which reduce the overall process success rate.

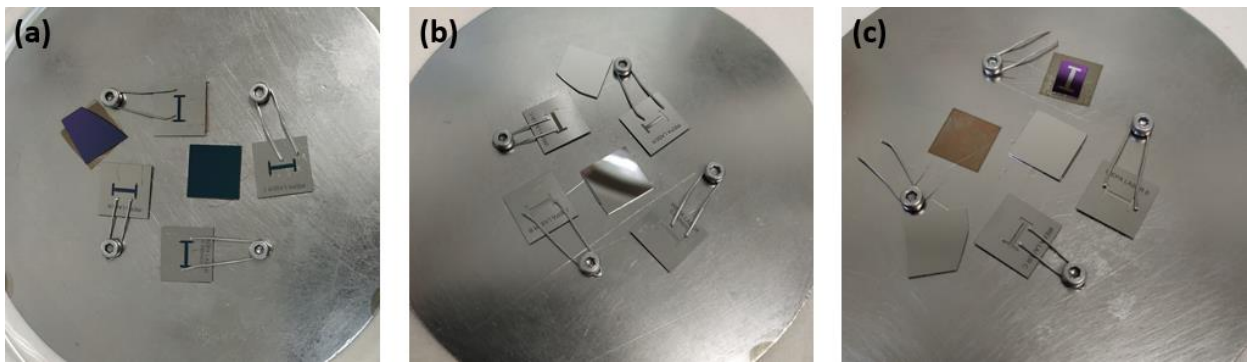

*Figure 8: Loose mask 1 implementation for lower electrode. (a) Before deposition. (b) After deposition. (c) After deposition without mask for 1 sample (see section 2.1).*

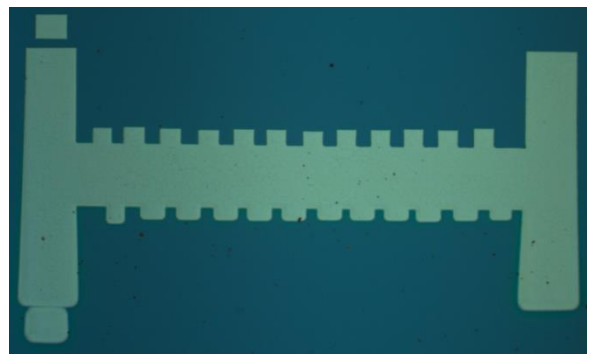

*Figure 9: Optical microscopic image of the lower electrode.*

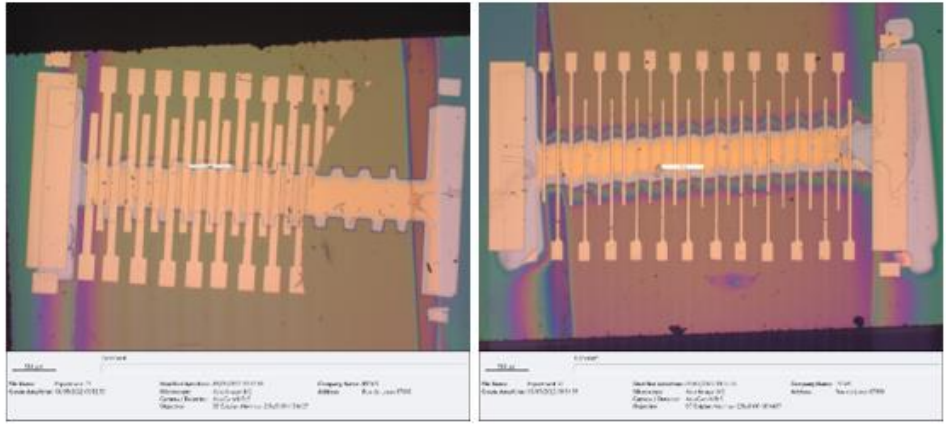

*Figure 10: Completed junction process using the Loose Mask implementation. These examples are among those with the most lateral/angular deviations from ideality.*

### 3.2.2. Auto-aligned Mask 1 Implementation

To address issues of alignment, to increase the number of junctions per deposition, and to deploy wedge technology (see Section 3.2.3), we conceived the solution schematized in Figure 11. It consists in a 3mm Cu plate with threaded holes that enables a 35x35mm Si substrate to be affixed to this plate using two clamps (seen horizontally) throughout the process. Each mask is then clamped onto this substrate holder using two other clamps (seen vertically). A closeup of the entire heterostructure patterned in the shape of a lower electrode is shown in Figure 12.

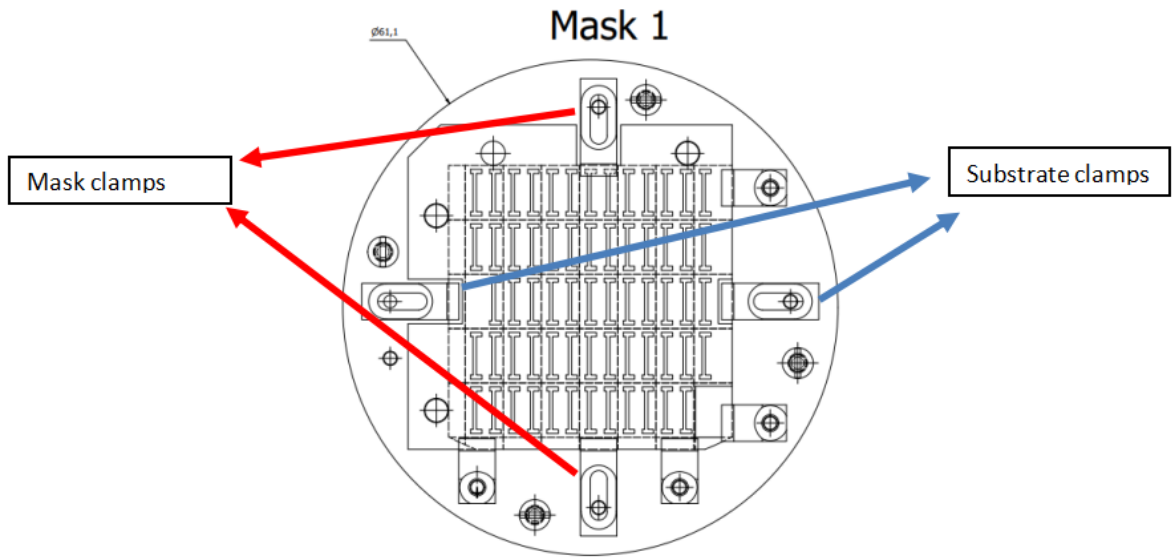

*Figure 11: Auto-aligned solution: Mask 1 for heterostructure deposition: schematic of the sample holder plate, sample and mask 1 held by clamps.*

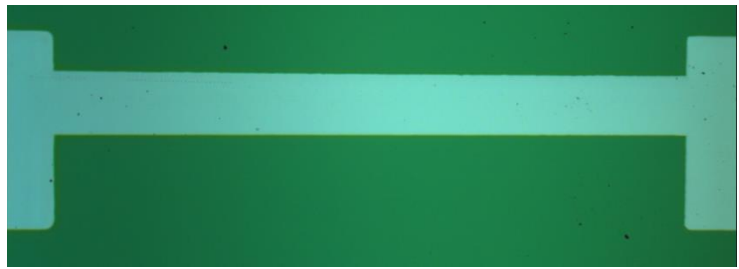

*Figure 12: Auto-aligned mask 1 implementation (see section 2.1): optical view of an entire stack deposited though mask 1, into the shape of the lower electrode.*

To obtain a repeatable alignment between the substrate and the process's three masks, the substrate holder features 3-4 pegs[1] that protrude from the plate upper surface. This allows the substrate to be butted up against the pegs, thereby pinning the substrate's position relative to the process patterns. Then, the mask is aligned by pushing it onto the pegs, thanks to holes for these pegs on the mask edges. Our experience has shown that, due to mechanical imperfections between the pegs and the mask, the mask can buckle and therefore should be pinned down with additional clamps (see side clamps in Figure 11). Examples of a deposition without and with these additional side clamps are shown in Figure 13. As a note, after first wedging separate pegs into the holder, we finally opted for a one-piece machining approach to improve mechanical accuracy between the substrate holder and masks, so as to reduce buckling.

---

[1] Having 4 pegs makes the problem hyperstatic, so one was removed.

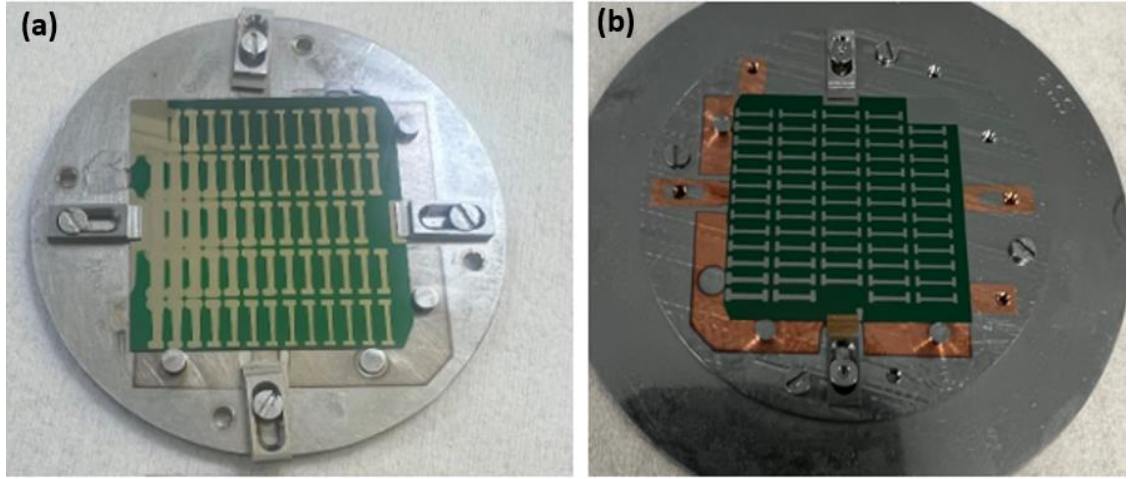

*Figure 13: Images of a full stack deposition through auto-aligned Mask 1 (a) before adding clamps, and (b) after adding clamps (see section 2.1). The remaining imperfections are the result of mechanical imperfections between the mask and sample holder pegs.*

### 3.2.3. Wedge molecular deposition

The large area of the 35x35mm Si substrate enables in-growth variations on the stack composition. In the context of energy harvesting/restitution samples[25], this is particularly useful: the $C_{60}$ layers can be varied in order to tune the tunnel coupling between ferromagnetic electrodes and paramagnetic centers, and thus the effective magnetic field that is electronically effective on the latter; and the CoPc thickness can be varied in order to adjust the length of the spin chain in the nanotransport path. To do so, a x-y motorized rectangular shutter is moved across the sample, or positioned so as to shadow certain sections.

This enables a wide range of compositional variations on a base stack structure. An example of these capabilities is schematized in Figure 14.

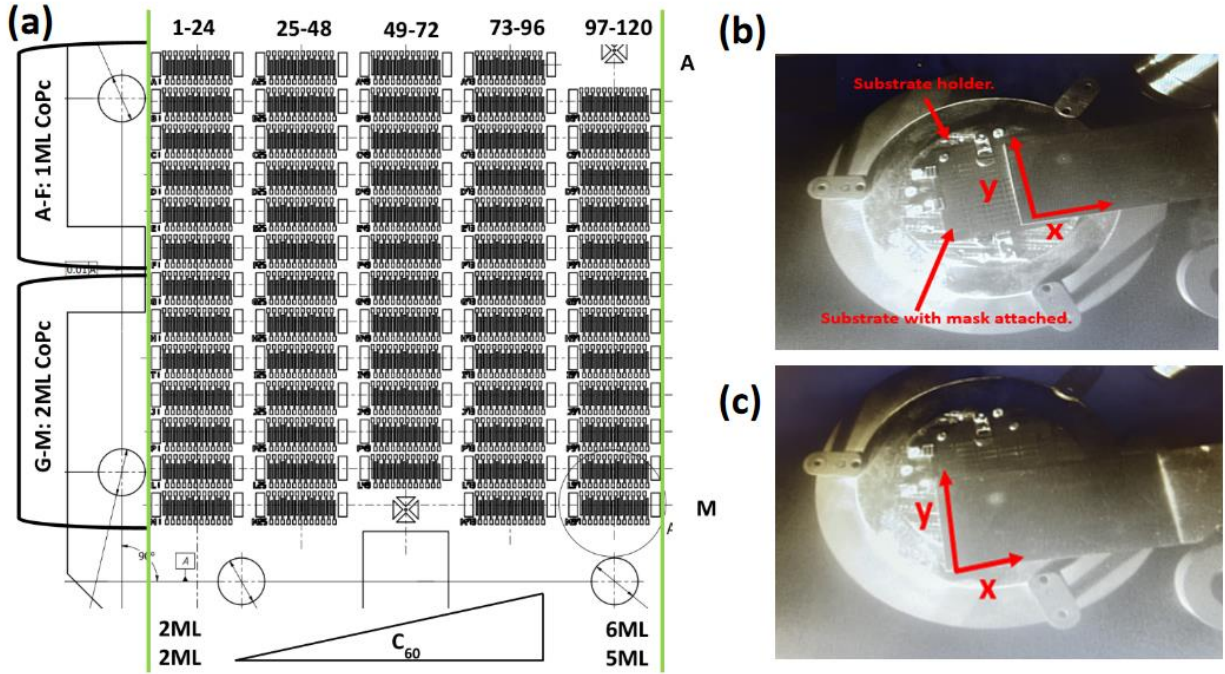

*Figure 14: Wedge deposition of a molecular spintronic heterostructure. (a) Schematic of the spatial position of a Si/SiOx//Cr(5nm)/Fe(45nm)/$C_{60}$(2ML)/$C_{60}$(2-6ML)/CoPc(1-2ML)/$C_{60}$(2ML)/$C_{60}$(2-5ML)/Fe(15nm)/Cr(100nm) stack. Mask 3 is depicted to show the effective junction density along the wedges. The green lines indicate the edges of the wedge. The left-hand brackets indicates the zones with 1 or 2ML of CoPc. (b-c) Pictures of the experimental setup for wedge deposition of a molecular layer. The pictures show a rectangular shutter translating along the X-Y plane between the sample and the molecular sources.*

# 4. Synthesizing Nanobeads

The heart of the process lies with utilizing nanobeads as the mask to define the vertical nanopillars. We chose $SiO_2$ as the nanobead material, figuring that its etching ratio relative to that of the metal layers to be etched would prove compatible. From the process summary of Section 2, one sees that a nanobead should not only remain in position once placed on the sample surface, but also be removable. Figuring that this level of control over adherence would require the tuning of the chemical properties at the capping layer (e.g. Cr., see Section 3.1), we not only purchased commercial nanobeads, but also synthesized them using the *surfactant mediated sol-gel synthesis technique*[44]. In this technique, we used tetraethyl orthosilicate (TEOS) as a precursor while cetyltrimethylammonium bromide (CTAB) was used as a cationic surfactant. Worthy to note is that the obtained nanobeads correspond to mesoporous silica nanoparticles having ca. 2.5 nm pore size however their main interest here is the possibility to finely control the nanobead diameter in the range 100-500 nm which is easier than the non-porous silica protocols. Especially the synthesis is performed at neutral pH (phosphate-buffered saline, PBS pH7) and it was shown that the growing addition of an alkanol co-solvent : propanetriol (PT) in the range 5-20% vol. allows to control the resulting nanobeads size from 100 to 300 nm. This effect was attributed to an increase of the hydrophobicity of the mixed solvents which seems to favor the condensation vs nucleation rates In this section, we will show how the co-solvent ratio (15 vs 12% vol. ratio) , the TEOS silica precursor addition mode (single or multistep), and the reflux time (12 vs 8h) determines the average size of the silica nanobeads.

Aiming for the synthesis of silica nanobeads of 300nm size[42], 1.07 gm of cetyltrimethylammonium bromide (CTAB) was completely dissolved in a mixture of 15% volume of propanetriol (44.1 ml)/phosphate-buffered saline (PBS, 250 ml, pH 7: 1.715 gm $KH_2PO_4$ and 0.29 gm NaOH) solution at $95^0$C under vigorous stirring using the magnetic bead. After attaining a clear, transparent, and homogeneous solution, the precursor TEOS of 4.5 ml was added dropwise in a single-step and let the reaction run for the next 8hrs. After continuous stirring for 8 hrs, the as-synthesized milk-white precipitate was collected by centrifugation process (8000 g/8 min) and then washed three times with ethanol (8000 g/8 min). To remove the surfactant (CTAB), the as-synthesized material was calcinated at $550^0$C for 6hrs.[42] To check the size distribution, a small concentration of the synthesized silica nanobead powder dispersed in the ethanol solution was drop cast on the Si substrate, followed by a natural drying process placed inside the scanning electron microscope (SEM)[42]. Figure 15 shows a typical SEM image and the size distribution of the synthesized silica nanobeads.

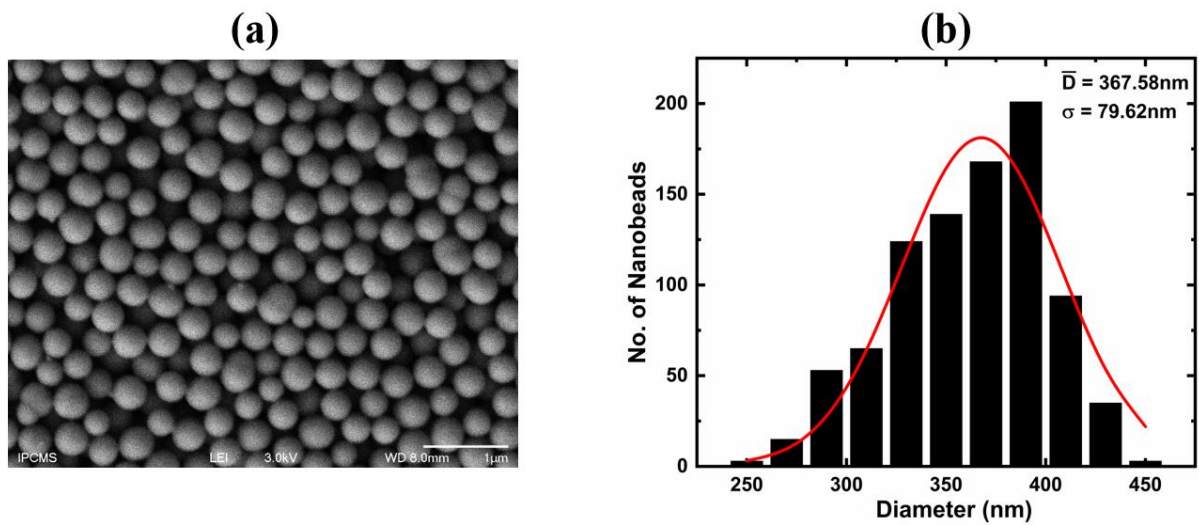

*Figure 15: Growth of silica nanobeads: first results. (a) SEM image of monodispersed silica nanobeads. (b) Statistics on the nanobead diameter. The average size of the synthesized silica nanobeads was calculated using normal distribution fitting for the extracted values from ImageJ software[45,46]. The average size ($\bar{D}$) of the synthesized silica nanobeads was found to be 367.58nm with a standard deviation (σ) of 79.62nm.[42]*

Using ImageJ software[45,46], the diameter of 900 random nanobeads was extracted. On fitting the normal distribution (see Figure 15 (b)), the average diameter ($\bar{D}$) was found to be 367.58nm with a standard variation (σ) of ±79.62 nm[42]. The synthesized silica nanobead thus exhibits a monodisperse morphology.

To reduce the nanobead diameter[42], the concentration of the surfactant solution was decreased from 15% to 12%. To implement the reaction with modified reaction parameters, 1.035 gm (for 10 mM concentration) of cetyltrimethylammonium bromide (CTAB) was completely dissolved in a mixture of 12% volume of propanetriol (PT 34.1 ml)/ phosphate-buffered saline (PBS, 250 ml, pH 7: 1.715 gm $KH_2PO_4$ and 0.29 gm NaOH) solution at 95$^0$C under vigorous stirring using the magnetic bead. After attaining a clear, transparent, and homogeneous solution, 4.5 ml of TEOS was added dropwise and the reaction was run for the next 8 hours. After continuous stirring for 8 hours, the as-synthesized milk-white precipitate was collected by centrifugation (8000 rcf/8 min) and then washed three times with ethanol (8000 rcf/8 min). To remove the surfactant (CTAB), the as-synthesized material was calcinated at 550$^0$C for 6 hours. Figure 16 shows the SEM image and the measured average size of the synthesized silica nanobeads. After decreasing the concentration of the surfactant (CTAB) solution from 15% to 12%, the average size ($\overline{D}$) of the silica nanobeads was decreased down to 201.29nm with the standard deviation (σ) of ±32.84nm[42]. This reduction in the average size of the silica nanobeads was more than what we aimed for (300 nm). Furthermore, from the SEM image[42] (compare Figure 16(a) and 16(a)), instead of the monodispersed distribution of the silica nanobeads, we observe some agglomeration.

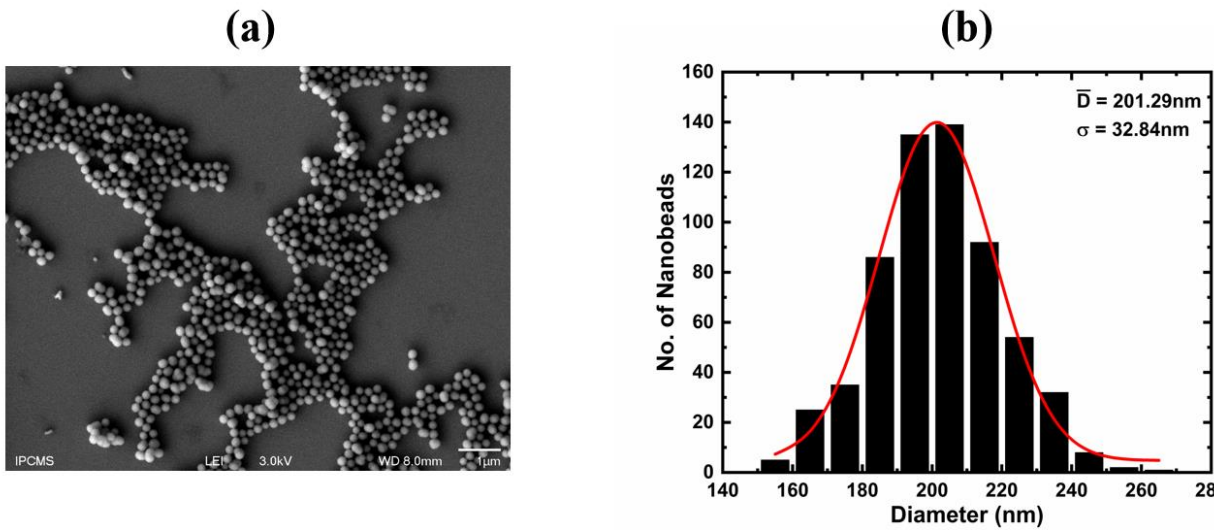

*Figure 16: Attempt to reduce the size of silica nanobeads. (a) SEM image indicates the formation of agglomerates of the silica nanobeads (b) Average size of the synthesized silica nanobeads was calculated using normal distribution fitting for the extracted values from ImageJ software[45,46]. The average size ($\overline{D}$) of the synthesized silica nanobeads was found to be 201.29nm with the standard deviation (σ) of 32.84nm.[42]*

In a final optimization toward the targeted 300nm diameter[42] (see process thickness considerations in Section 2.6), and to reduce agglomeration, we kept the concentration of the surfactant (CTAB) solution fixed i.e., 12% but modified the process of adding the precursor solution (TEOS) and the reflux time. The existing single step addition of the precursor solution (4.5 ml of TEOS) to the surfactant solution was replaced with a multistep addition process and the reflux time was increased from 8hrs to 12 hrs. The complete chemical reaction was performed in the following way: First, 1.035 gm (for 10 mM concentration) of cetyltrimethylammonium bromide (CTAB) was completely dissolved in a mixture of 12% volume of propanetriol (PT 34.1 ml)/ phosphate-buffered saline (PBS, 250 ml, pH 7: 1.715 gm $KH_2PO_4$ and 0.29 gm NaOH) solution at 95⁰C under vigorous stirring using the magnetic bead. After attaining a clear, and homogeneous solution, 750 µl of TEOS was added dropwise and the reaction was left to run for 2 hours. This dropwise addition of 750 µl of TEOS into the reaction was repeated 5 more times. The total 6 times addition of 750 µl of TEOS is thus equivalent to a 4.5 ml TEOS addition but in a multistep process with a total reflux time of 12hrs. Compared to the single addition, the multistep addition of TEOS provides better control on the growth of the silica nanobeads with the desired size of 300 nm. After continuous stirring for 12 hours, the as-synthesized milk-white precipitate was collected by centrifugation (8000 rcf/8 min) and then washed 3 times with ethanol (8000 rcf/8 min). To remove the surfactant (CTAB), the as-synthesized material was calcinated at 550⁰C for 6 hrs.[42] To protect the silica nanobeads from environmental humidity, the calcinated powder was stored in an airtight vial. Figure 17 (a) shows the improved morphology of the silica nanobeads. The average diameter ($\bar{D}$) measured for 300 random silica nanobeads using the ImageJ software[45,46] and fitted with a normal distribution curve was 297.40 nm with a standard deviation (σ) of ±38.24 nm[42] Figure 1717 (b). The average values of the silica nanobeads with the standard deviations are very close to 300nm aimed for the nanolithography process[42].

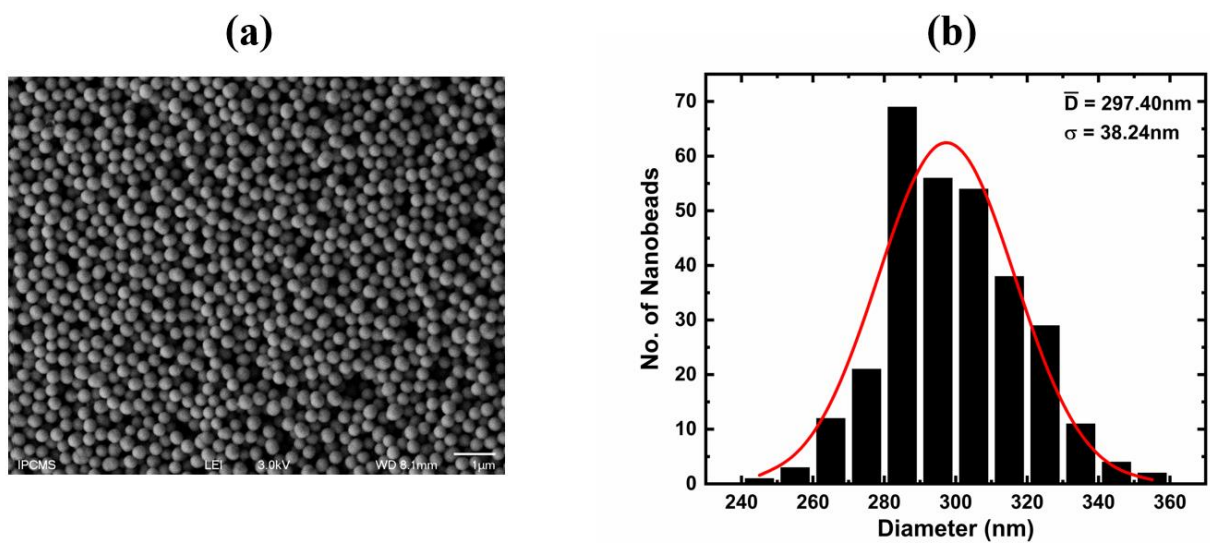

*Figure 17: Growth of silica nanobeads: final result. (a) SEM image shows monodispersed silica nanobeads, (b) Average size of the synthesized silica nanobeads calculated using normal distribution fitting for the extracted values from ImageJ software[45,46]. The average size ($\bar{D}$) of the synthesized silica nanobeads was found to be 297.40nm with the standard deviation (σ) of 38.24nm.[42]*

# 5. Nanobead masking

As shown in Figure 2, there should ideally be one nanobead at the intersection between the top and bottom macroelectrodes defined by process steps 1 and 5 (see Section 2). Given the overall requirement that the process not use solvents/liquids, we initially undertook various dry approaches: dry dispersal (Section 5.1) and dry positioning (Section 5.2). Then, to reduce randomness and increase the process success rate, we turned to microfluidic micrometric positioning (Section 5.3) after confirming that the solvent picodroplets were not affecting the heterostructure thanks to its capping layer.

### 5.1. Dry Dispersal

As a first approach used by our research group to deposit the silica nanobeads onto the stack's capping layer, we adapted[47] earlier research[48]. The pre-processed silica nanobeads in dry powder form are kept in a petri dish (silica nanobead reservoir) and the sample is maintained above the petri dish with the sample surface facing downwards toward the petri dish. This whole setup is then placed in an ultrasonication bath, and an ultrasound pulse is applied. The energy causes the silica nanobeads to jump onto the sample surface. By optimizing the distance between the sample and the nanobead reservoir, the ultrasonic power, and the ultrasonic pulse duration, the number and areal density of nanobeads deposited on top of the sample surface (capping layer) can be qualitatively controlled. Using 45 µm polystyrene beads as a test, a representative optical picture of the capped hybrid trilayer stack coated with dispersed beads is shown in Figure 18. For micrometric bead sizes, deposition is done by gently tapping a micro spatula above the sample surface. The beads are large enough to be visible to the naked eye, allowing real-time evaluation of their distribution and density. If necessary, the surface can be cleaned with a nitrogen gun and the process repeated.

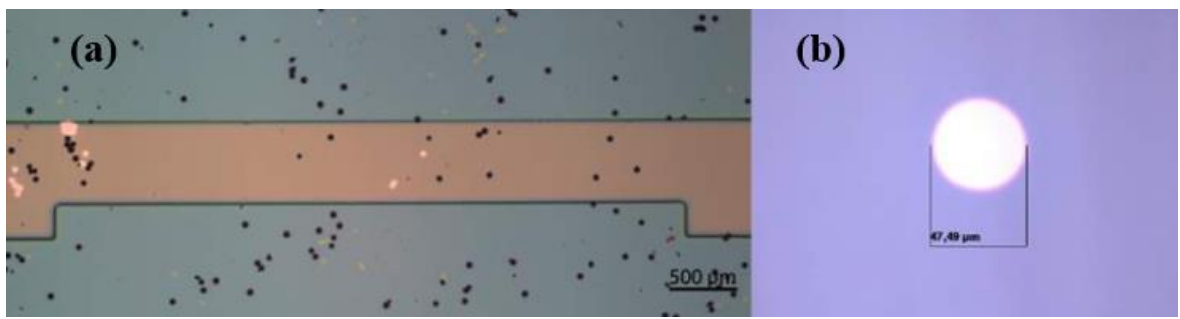

*Figure 18: Dry nanobead dispersal. (a) A representative microscopic image of the dry bead process performed on a capped hybrid trilayer stack with (b) polystyrene microbeads of the size of approx. 45µm. The figure is taken from ref[47]*

After optimizing the deposition parameters, the refined dry deposition technique was used to fabricate the hybrid nanopillars using 500nm $SiO_2$ beads. For nanometric beads, only clusters are visible to the eye, making control more challenging. Tapping the spatula typically results in excess deposition. Instead, the beads can be placed in a Petri dish floating in an ultrasonic bath with either pulsed or continuous power modes. The ultrasound causes them to jump, and by holding the sample with tweezers 0.5–2 cm above the dish (without direct contact), a small amount can be deposited. A brief exposure of less than <1 s is usually sufficient. Parameters such as bead size, weight, and agglomeration tendency significantly influence the result. Both pulsed and continuous ultrasonic modes seem effective; choice depends on operator preference.

Although the dry deposition technique produces working nanopillars[26], there is no control over the number of the nanobeads and their precise location on the top surface (capping layer) of the stack, which can result in parallel junctions underneath a top electrode (see section 2.5). To control the number of silica nanobeads during the deposition process, the sample was covered with two copper TEM grids with a mesh placed in a 45degree crossed geometry, which act as a sieve. Although this modification avoids bigger agglomerates of silica nanobeads from reaching the sample surface, this does not reduce the number of silica nanobeads adsorbed on the sample surface significantly. The second issue i.e., the positioning silica nanobeads on the sample surface at the desired location, however, could not be resolved using this dry deposition scheme.

## 5.2. Dry positioning

To resolve the underlying issues with the previously mentioned dry dispersal technique in section 5.1, a nanomanipulation scheme was implemented using an atomic force microscope (AFM). In this scheme the silica nanobeads are deposited on the top surface (capping layer) of the heterostructure stack.

### 5.2.1. Conceptual explanation

The AFM is a surface characterization tool widely used to image the topography of the sample by measuring the atomic force between the sample surface and the sharp AFM-tip. However, using the in-built *dynamic mode* popularly known as *tapping mode* together with turning feedback mechanism off, one can manipulate the nano species (such as nanoparticles in our case) adsorbed on the stack surface, thereby controlling their numbers and spatial distribution to enable the precise formation of the nanopillar(s) followed by the dry etching step, at desired location on a given stack. The AFM-nanomanipulation process is schematically presented in Figure 19.

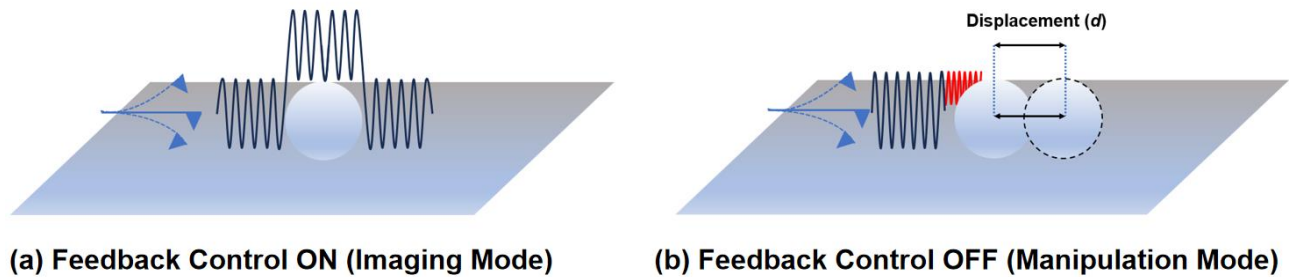

*Figure 19: Schematic representation of the AFM nanomanipulation achieved in tapping mode. During the imaging process, the feedback remains ON, while the lateral manipulation/displacement (d) is achieved by turning OFF the feedback control. The figure is adapted from Ref [49].*

### 5.2.2. AFM-nano manipulation results

In this AFM-based nanomanipulation scheme, we have used the following two strategies: i) wiping out the undesired silica nanobeads from the sample's surface by selecting a large manipulation area of 10µm, followed by ii) manipulating a single silica nanobead on the surface.

To wipe away undesired silica nano beads from the surface requires moving the nanobeads.

Figure 20 shows the displacement of the adsorbed silica nanobeads on top of the sample's surface (capping layer). The brown background represents the sample surface, while the adsorbed nanobeads are represented by the white spherical shape-like objects. With proper tuning of the feedback parameters, including the amplitude setpoint, the silica nanobeads in panel

Figure 20(a) were shifted towards the right direction (see panel b). By continuing the manipulation process over a large area of 10µm over a span of several hours, it is possible to move most of the silica nanobeads towards the boundary on the right-side panel

Figure 2021 (c), and retain and fewer silica nanobeads in the middle portion. Panel

Figure 2021 (d) shows the remaining silica nanobeads adsorbed on the sample's surface.

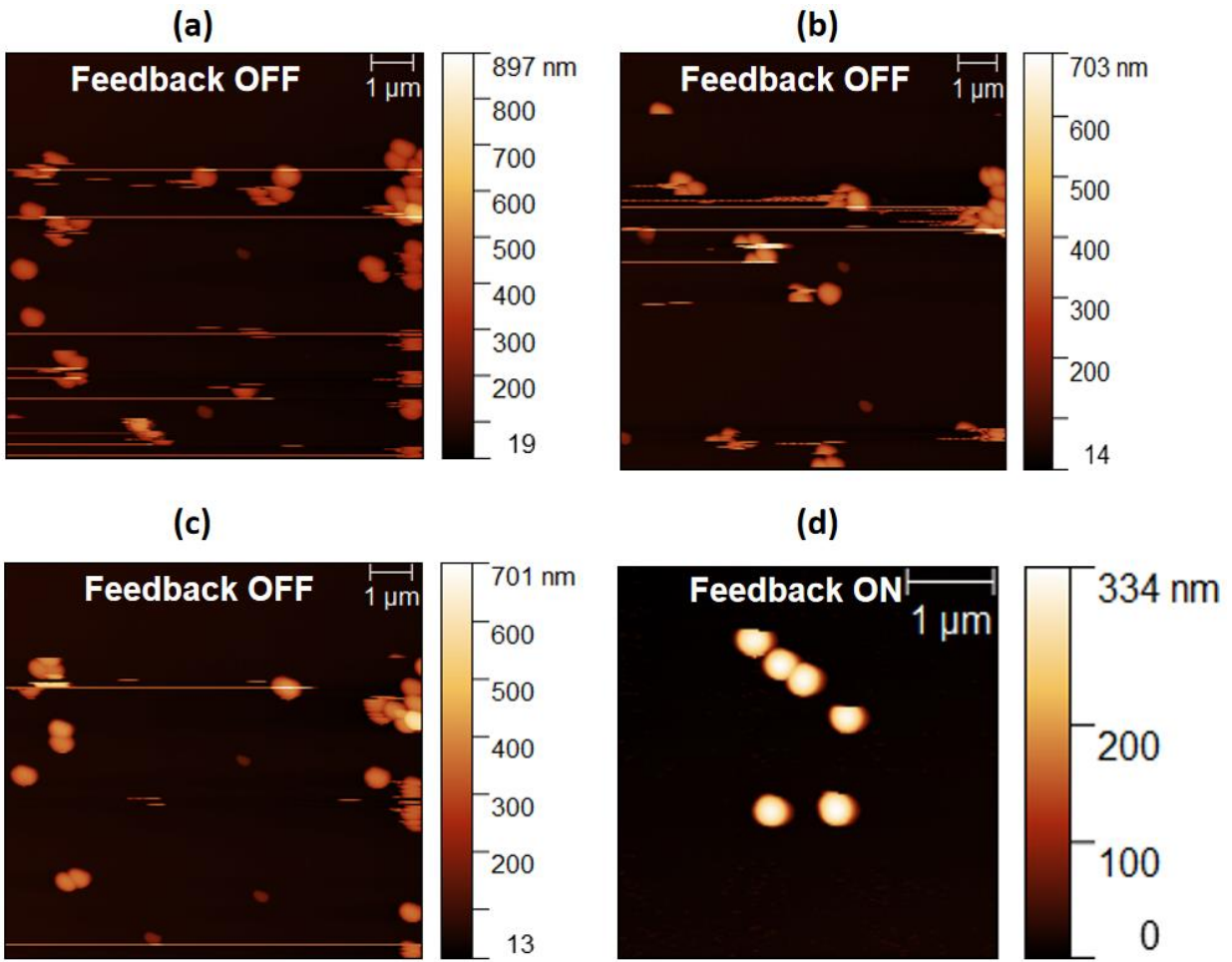

*Figure 20:. AFM images collected and plotted[50] on the panel (d) after a series of lateral manipulations performed and on silica nanobeads of ~300 nm size loosely bound to the top surface (capping layer) of the capped hybrid trilayer stack shown by the panel (a, b, c).[42]*

Comparing the initial distribution of adsorbed silica nanobeads (

Figure 20(a)) to the final situation panel (

Figure 20(d)) , a significant reduction in the number of adsorbed nanobeads has been observed. On applying the AFM-nanomanipulation techniques discussed in the previous subsection on a reduced scan/manipulation area of approximately 4x4µm² on the sample surface, we were able to move an adsorbed silica nanobead up to 1.45µm along the in-plane (*x-y*) direction.

Figure 21 shows the AFM images of the adsorbed silica nanobeads before and after the AFM-nanomanipulation scheme was implemented.

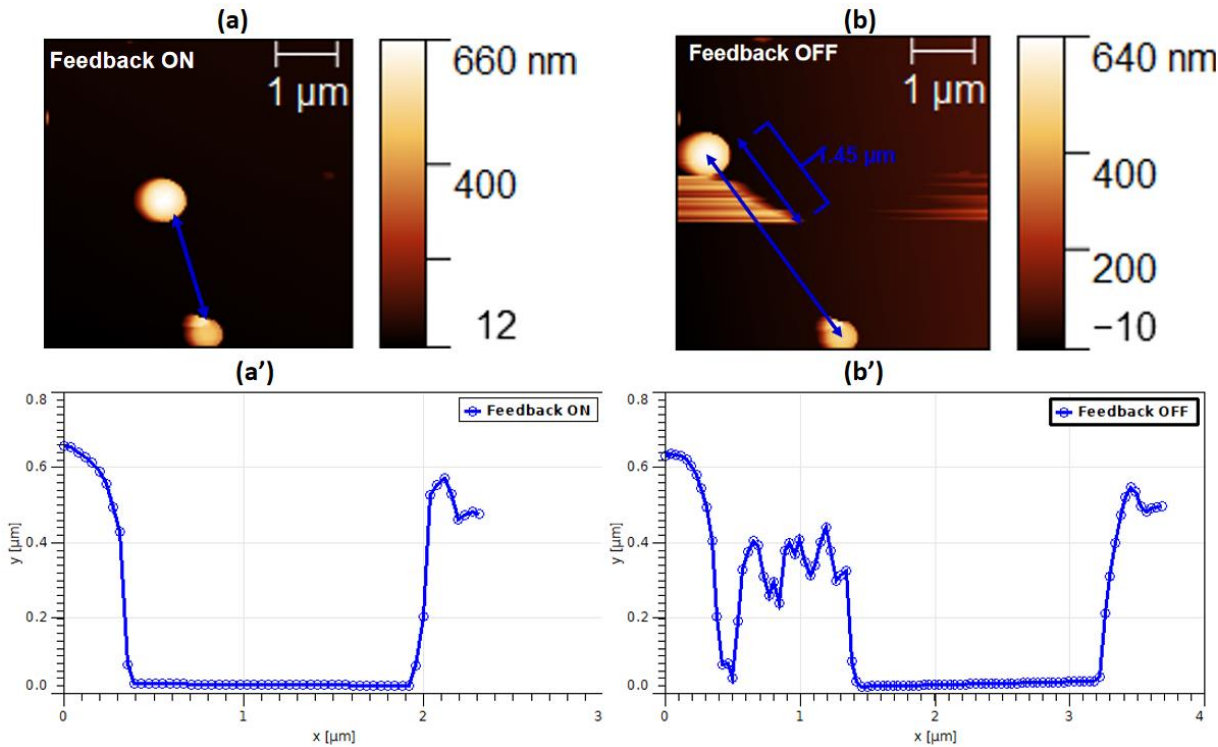

*Figure 21: AFM image of a silica nanobead (a) during the imaging process, and (b) after the lateral manipulation, and calculated distance between the two nanobeads (a') during the imaging process and (b') after the lateral manipulation. The lateral manipulation displacement (d) of 1.45µm thus corresponds to approximately 5 times the size of the silica nanobead of 300nm size. [42]*

Using Gwyddion software[50], the distance between the two adsorbed silica nanobeads was calculated for both before and after the nanomanipulation. The distance before the nanomanipulation was found 1.55µm, while after performing the nanomanipulation it was found to be 3.00µm. This displacement of 1.45µm achieved after the AFM-nanomanipulation corresponds to approximately 5 times the size of a 300nm silica nanobead. Although the AFM-nanomanipulation technique was found to be very effective for laterally displacing the silica nanobeads adsorbed on the surface (capping layer) of a capped hybrid trilayer stack, for a stack with a comparatively large surface area, this is a very time-consuming process.

This difficulty for samples with large surface areas arises due to the smaller scanning size (max ~10μm for good quality scans) and larger scan time (approx. 25-40 minutes for each scan) taken by AFM during both imaging and manipulation processes. To overcome this issue, one option is a nanomanipulation scheme done along the z-direction. In this process, AFM-tip would be used to pick a single silica nanobead from the reservoir and could be placed over the desired location on the top surface (capping layer) of the capped hybrid trilayer stack. First attempts were made but this approach was abandoned due to time constraints. This approach, when coupled with machine learning, should be able to work but would remain time-consuming.

## 5.3. Microfluidic Micrometric positioning

The work of Section 5.2 revealed that a dry positioning approach is extremely time consuming, and would require automation on the AFM-driven manipulation that is not available commercially, to the best of our knowledge. We therefore considered a microfluidic approach, in which the nanobead is positioned at the intersectional area between bottom and top electrodes using a microdroplet carrier. This approach assumes that the heterostructure's capping layer strategy is resilient against this local exposure to the microfluidic ink's solvents. Our first attempts (Section 5.3.1) using a manual approach confirmed this technological assumption, and paved the way for a more automated matrix implementation (Section 5.3.2) that is the process's present state-of-the-art at the time of this writing.

### 5.3.1. First attempts

This Section describes our first attempts to position silica nanobeads by printing microdroplets of an ink containing these nanobeads.

#### 5.3.1.1. Equipment

To print the microdroplets, a fine micropipette was attached to a commercially available microinjector. The selection of the deposition location is done using a manual/semi-automatic manipulator and movable sample stage. The in-house developed setup of the microfluidic pen lithography is shown in Figure 2223(a), and its operation is schematized in Figure 2223(b). The manual/semi-automatic manipulator holds the micropipette while the sample stage holds the capped hybrid trilayer stack on which the deposition of microdroplets containing the silica nanobeads is to be performed. Both manual/semi-automatic manipulator and sample stage have the translational degree of freedom in the *x-y* plane and the rotational degree of freedom of $360^0$ at *y-z* and *x-y* planes, respectively. The entire process of the silica nanobead deposition is carried out under the optical microscope/digital camera with a sufficiently large working distance.

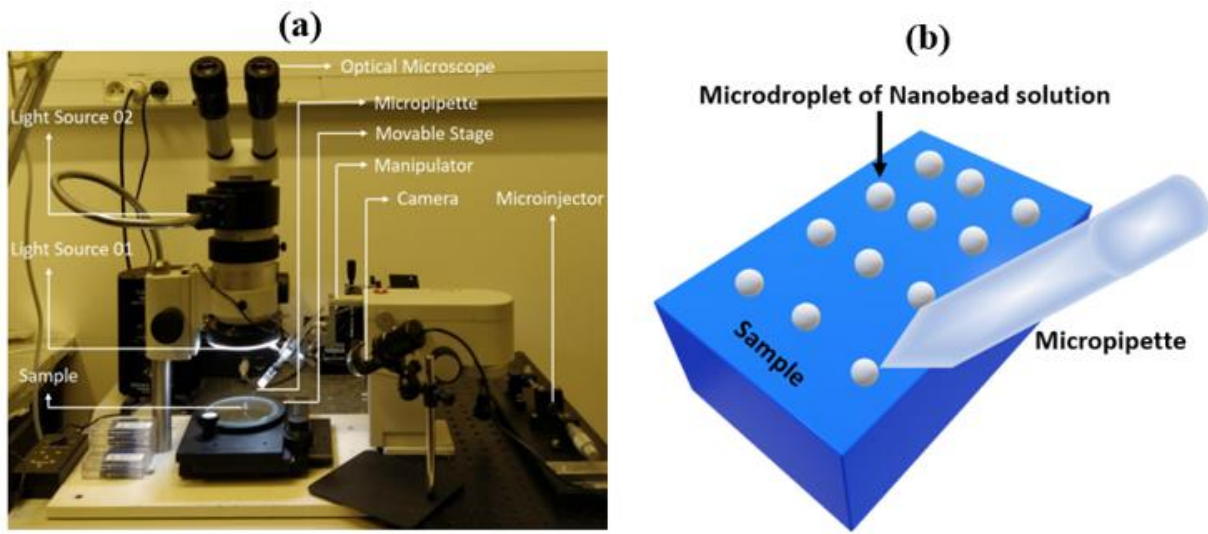

*Figure 22: In-house developed 'microfluidic pen lithography' setup. (a) The in-house developed setup of microfluidic pen lithography, and (b) Schematic presentation of the working principle of microfluidic pen lithography.*[42]

This ejected droplet of a few μm (3-10 μm) diameter (or picolitre volume) from the micropipette is placed on the desired location on the top surface (capping layer) of the heterostructure and dried at a nominal applied temperature (~50$^0$C) for 1 hour maximum.

### 5.3.1.2. Ink Preparation

Given the equipment's ability to easily prepare microdroplets with a 10 μm diameter, and the desired requirement that each 10 micron microdroplet contain around 1 nanobead, we prepared ink in the following way. Silica nanobeads are dispersed in a 5:1 mixture of ethanol and glycerol. Ethanol acts as a carrier to flow within the micropipette as ethanol has a low viscosity of 1.074 mPa.s at room temperature. To avoid the spilling of droplets on the sample surface, the droplet's surface tension needs to be increased. To do so, glycerol whose viscosity is 1.412 Pa.s at room temperature is added. To prevent bead fusing we have to control the concentration of beads in the ink. First attempts were made with an Ethanol-Glycerol (5:1) ratio. To make the ethanol-glycerol combination in the appropriate (5:1) ratio, 1.00 mL of absolute ethanol was measured with a micropipette and transferred to a clean glass vial. Glycerol was first measured by mass to assure formulation precision. A 254 mg glycerol sample was weighed with an analytical balance. Given the known density of glycerol (≈1.26 g/mL at room temperature), this mass was converted into its volume. The conversion was necessary because handling and mixing liquids is more convenient in volumetric terms rather than mass. The conversion yielded around 0.20 mL of glycerol, which was then added to the ethanol.

After preparing the ethanol–glycerol (5:1) mixture, 2 mg of 500 nm-diameter silica monodisperse, non-porous beads were accurately weighed using a microbalance and dispersed into 1.20 mL of solution Ethanol-Glycerol (5:1), yielding a final concentration of 1.66 mg/mL. This concentration was determined experimentally: initial trials with higher concentrations (≈4 mg/mL) resulted in an excessive bead density per 10 μm droplet during printing. By progressively lowering the bead mass, and hence the ink concentration, to the range of 1.5–2 mg/mL, the bead count per droplet stabilized to the desired level (typically 0-3 three per droplet).

To further reduce the junction lateral size, 300 nm silica beads were employed using the same preparation method. In this case, the mass of beads was reduced to 0.20 mg, which was dispersed into 1.20 mL of the ethanol–glycerol mixture, yielding a concentration of 0.17 mg/mL. This lower concentration was specifically chosen to obtain an average of 1–3 beads per 10 µm droplet.

### 5.3.1.3. Results

A representative optical image of the capped hybrid trilayer stack on which this lithography technique has been performed is shown in Figure 23. In Figure 23 (a), the area with dark gray color represents the Si/SiOx substrate while the C type structure with white color is the heterostructure stack (see Section 2.1). The black dots that span the C-type structure are the micron-sized droplets with picolitre volume containing silica nanobeads SEM imaging reveals that a nanobead can be present in the area of the microdroplet. (see Figure 2324(b)).

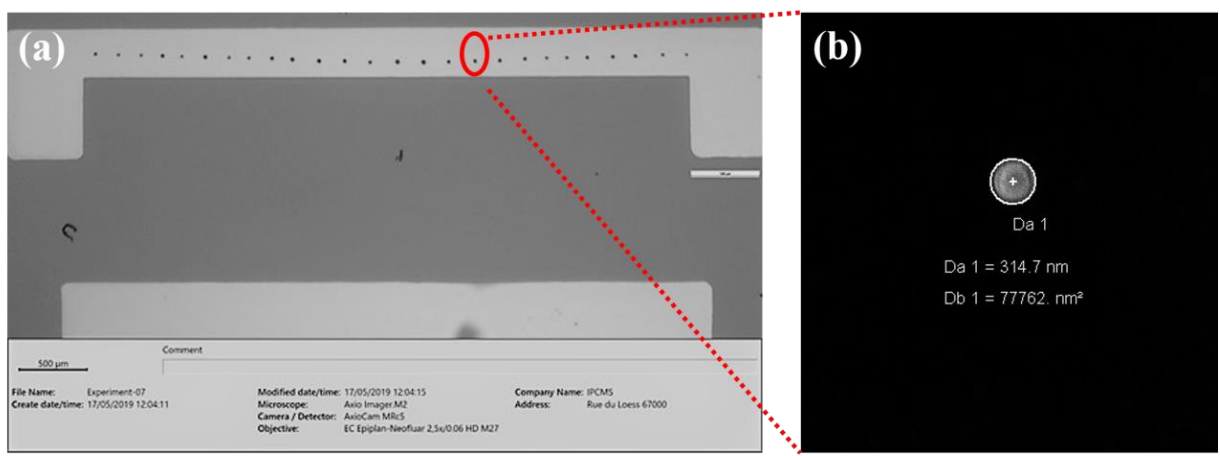

*Figure 23: First experiments on nanobead positioning using 'microfluidic pen lithography'. (a) Optical image of the capped trilayer hybrid stack after printing. In the image, the dark gray surface represents the Si/SiOx substrate while the lighter gray C-shaped structure is the heterostructure stack pre-patterned into a lower electrode using shadow-mask disposition. The black dots are the micron size droplets containing silica nanobead of ~300nm size. (b) SEM image of the nanobead left after the evaporation of the microdroplet.[42]*

### 5.3.2. Matrix implementation

The results of the previous Section demonstrate that it is possible to position nanobeads onto a sample with a precision that is acceptable for our process. Since one means to achieve the needed[2] increase in reproducibility of transport results is to study more junctions, a concerted effort along the technological chain was undertaken, with an implementation using 63 lower electrodes, each with 24 junctions (see Section 3.2.2). Printing 1512 nanobeads thus requires a matrix approach, which is described hereafter.

---

[2] This research line, and its results, represent a departure from model junctions, assembled using a scanning tunneling microscope or using lateral contacts. In those junctions, the contact is ephemeral as the junction is broken and remade 100s of times, and measurements occur upon each formation. This advantageously introduces the notion of statistical reproducibility, even though these approaches have little industrialization path opportunities. In contrast, this paper's approach is to assemble solid-state molecular vertical junctions that can potentially be used in applications. Our first results[26] using this process led after multiple review rounds to a demotion from consideration in Nature Materials because only 4 working junctions were reported.

### 5.3.2.1. Equipment

To achieve matrix printing of microdroplets containing a calibrated number of nanobeads, we acquired and installed on the STNano technological platform a Nano eNabler™ system. This commercial micropositioner leverages attoliter- to femtoliter-scale liquid dispensing technology, and allows for real-time printing process monitoring, using a XY stage platform with a 50x50 mm working area and 20nm positioning resolution. The ink is placed within a Surface Patterning Tool (SPT), a silicon-based microcantilever integrated with a microfluidic channel. Its design/geometry determine spot sizes/deposition volumes.

Printing can be done either by preset z-motion of the tip onto the sample surface, or a laser monitoring system can detect the mechanical bending of the cantilever upon touchdown. Software enables the repeated printing and sample displacement, according to a preloaded script, to achieve the desired matrix of microdroplets in accordance with the process-determined junction locations.

To prepare a SPT, the first step is to ozone-clean it (40 minutes in the provided cleaner), and then mount it on the SPT Holder using a Bio Force SPT adhesive pad or a piece of permanent double-sided stick tape. The next step is to load the ink. Ink concentration depends on the need which means the required number of beads in a droplet and the diameter of a droplet which depends on multiple factors such as type of SPT tip, surface roughness and humidity. See section 5.3.2.2 for details.

Once the SPT is secure, use a pipette to dispense approximately 0.5 µl straight down into the reservoir. The SPT and SPT Holder may be placed on the sample platform to allow high-magnification observation of the loading progress. As liquid/Ink fills the channel and flows toward the end of the cantilever, the channel will begin to disappear. Once you are satisfied with the required amount of ink the holder can be mounted on the micropositioner head.

### 5.3.2.2. Ink Preparation

The same basic tenets described in Section 5.3.1.2 are reapplied here, noting that the micropositioner can control humidity during printing. It is then important to obtain a reasonably reproducible microdroplet diameter across the entire wafer. To control bead adhesion, we have to be careful with the ethanol-to-glycerol ratio. if we increase the concentration of ethanol the ink will get dry inside the SPT tip and it won't be able to print. If we increase the concentration of glycerol the droplet will spread. We have tried multiple ink concentrations Ethanol-Glycerol (5:1), Ethanol-Glycerol (7:1), Ethanol-Glycerol (10:1) and Ethanol-Glycerol (5:2). Among these only Ethanol-Glycerol (5:1) ratio yielded satisfactory results (see section 5.3.1.2). With this ratio and control over other parameters, we got the desired number of beads as shown in Figure 24.

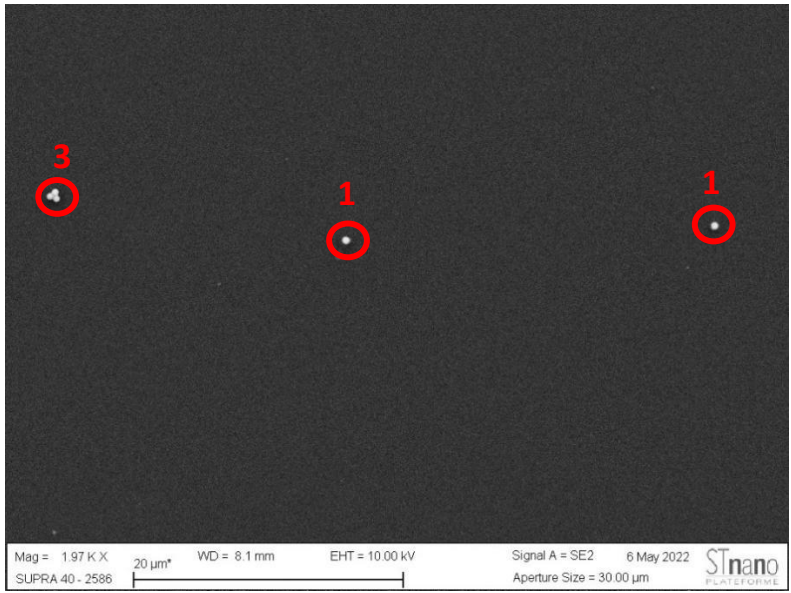

Figure 24: SEM image of printed array and the required number of beads shown in red circles. Statistics of the number of nanobeads per spot are given in section 5.3.2.3.

Environmental control is a critical feature of the micro positioner, with a fully enclosed chamber that manages humidity levels between 25% and 80% relative humidity (RH). This is achieved through a computer-controlled system that balances dry inert gas, such as nitrogen, and humidified air. Precise control over humidity is essential to ensure proper material transfer and feature size regulation. Insufficient humidity can hinder the transfer, while excessive humidity can lead to uncontrolled spreading of deposited materials.

### 5.3.2.3. Results

In principle, the micropositioner can be programmed to print an entire array of 1512 microdroplets to execute nanobead masking across the sample in one operation. However, while the tool was factory-optimized along the x direction with a deviation from level lower than 5 microns overs 35mm, this deviation increases to 25 microns over 35mm in the y direction. This means that either laser-feedback is implemented to auto-correct, or that the printing be broken up into substeps for which the actual z height of the SPT tool is adjusted. It is the latter approach that we are using at the time of this writing.

Furthermore, due to masking issues that were only lately corrected, the masking of the heterostructure during Step 1 could lead to ill-defined features (e.g. the corner of a lower electrode pad) that are then used to position the microdroplet array onto the entire heterostructure pattern. This can lead to a misalignment between microdroplet and the target surface area of the lower electrode pattern. Thus, in what follows[3], a simplified protocol was used: rather than print one microdroplet within the sample area that should contain the overlap between the junction's macroelectrodes, an array of microdroplets was printed. With an 85 μm top metallization electrode pitch and up to 5 μm positional error due to a 10 μm diameter microdroplet, we chose a pitch of 40 μm for our array along the lower electrode. A photo of a heterostructure with one such array being printed by the micro positioner is shown in Figure 25.

---

[3] To be complete, we are also utilizing the one-droplet-per-junction approach in day-to-day research.

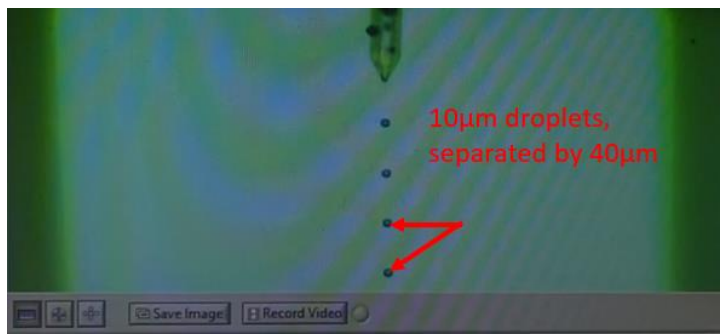

*Figure 25: (10μm) Droplets printed by micro positioner with a distance of 40 μm between them.*

The effective nanobead count per junction depends on different parameters such as ink concentration, the type of SPT tip, temperature/humidity conditions, the microdroplet diameter, etc. As seen in Figure 26 , this count was initially too high. The ink concentration was 4mg/ml, SPT type was 60 R, Humidity was 50% and the average droplet diameter is 13-15 um. By optimizing the parameters, we controlled the required number of nano beads per junction which is less than 3 as shown in Figure 27.

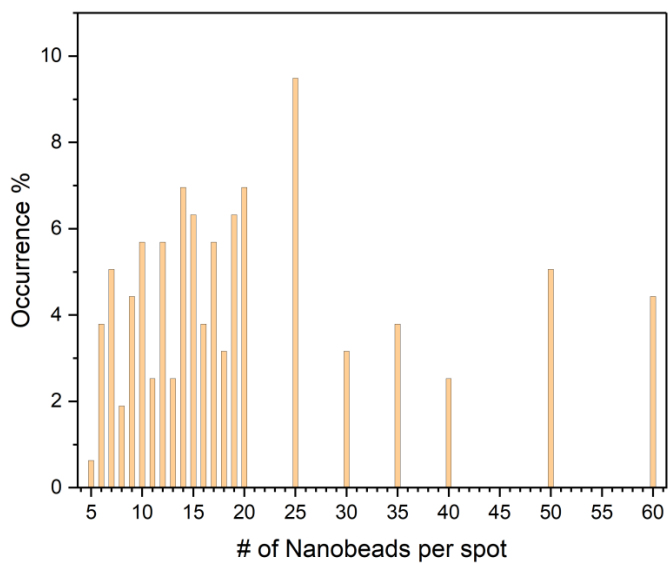

*Figure 26: Statistics on the nanobead count of a 158-member array of microdroplets that were printed before optimizing parameters.*

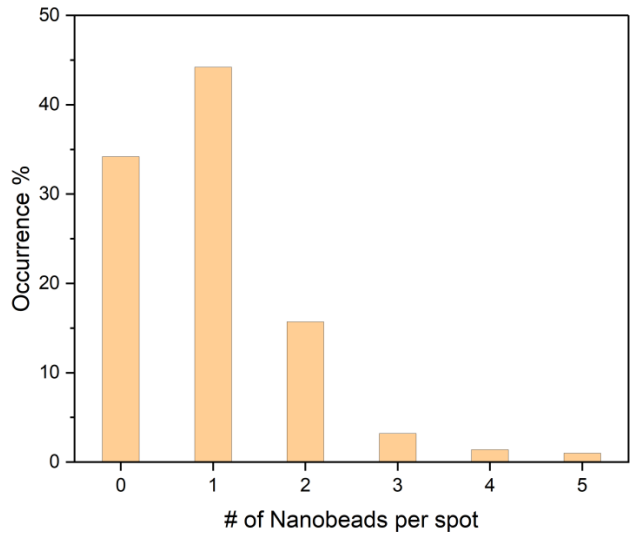

*Figure 27: Statistics of a 280-member array printed after optimization using the micropositioner so that almost half of the spots have one nanobead.*

As a note, it is at this stage of the process that we tested the ability to remove the nanobeads. Indeed, while this removal effectively takes place after dielectric encapsulation, the resulting insulating surface prevents any SEM observation of nanobeads that would have not been removed. Tests were therefore undertaken on samples in a state prior to etching, and after a simulated etching but before dielectric coating. This is covered in Section 7.2, which also discusses the impact of SEM observation on the ability to remove the nanobeads.

# 6. Ion Beam Etching and Encapsulation

Referring to Section 2.3, once the nanobeads are positioned on the heterostructures, the sample is etched using a neutralized Ar ion etching beam, and then encapsulated. This Section details the various steps required to achieve this result.

## 6.1. Mask 2: protecting lower electrode pads from dielectric encapsulation

As with oxide junctions[51,52], this vertical molecular nanojunction process deploys a strategy of achieving an electrical contact to the lower electrode across the entire heterostructure stack. This strategy integrates the expected presence of catastrophic defects in a $mm^2$ area of the heterostructure, such that an Ohmic contact will be obtained. If, in addition to the material under the nanobeads, the surface area to be used to achieve the lower electrode contact is also masked prior to the etching step, then the sample can, without breaking vacuum, undergo dielectric encapsulation. This in principle removes the need for an additional process step (e.g. reactive ion etching) to open a dielectric window to achieve this Ohmic contact onto the lower electrode.

### 6.1.1. Loose Mask 2

In the loose mask iteration of the process, Kapton tape is used to cover the lower electrode pads. Then the sample is etched, and then brought back to ambient air conditions, sample is then placed in the dielectric deposition chamber. This use of Kapton tape to achieve masking can result in a soft dielectric profile that is optically visible in (see Figure 28).

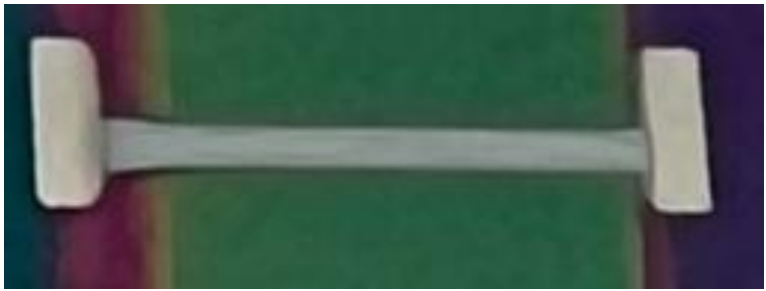

*Figure 28: Loose mask 2 implementation of dielectric encapsulation: optical image of a cell. Optical fringes close to the lower electrode contact are the result of a gradient in dielectric thickness due to imperfect masking using Kapton tape.*

### 6.1.2. Auto-aligned Mask 2

In the auto-aligned process, Mask 2 as shown in Figure 29 is used to protect the pads of the lower electrode from dielectric encapsulation (see section 2.3). Overall, the result is better masking (see Figure 30).

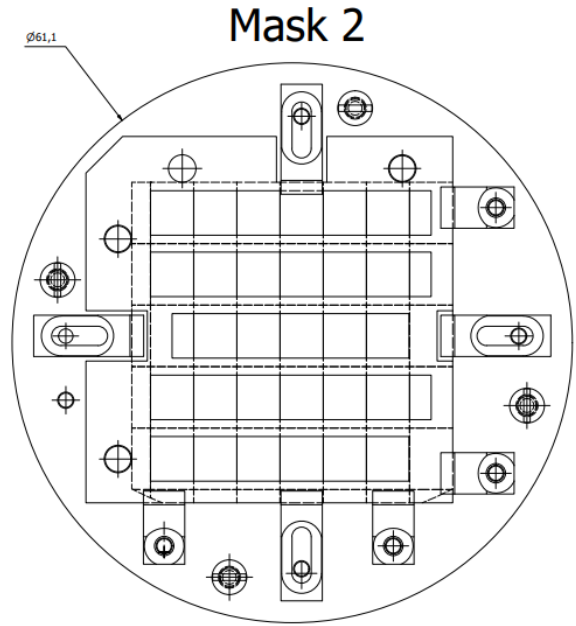

*Figure 29: Auto-aligned process: schematic of sample plate, sample and Mask 2 before etching and dielectric encapsulation (see section 2.3).*

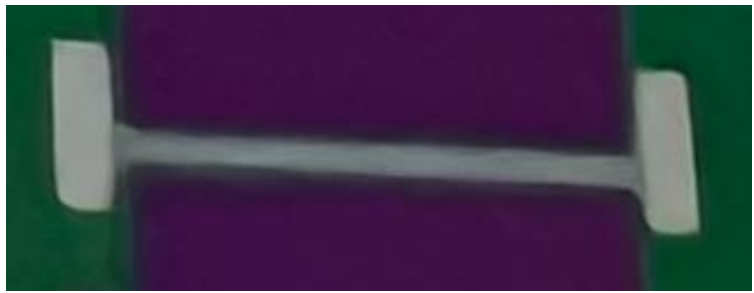

*Figure 30: Auto-aligned Mask 2: photo of sample cell after IBE and encapsulation. The mostly homogeneous lateral profile of the lower electrode suggests a reduction in thickness variation due to masking issues during Step 1.*

### 6.2. Separate etching and dielectric encapsulation tools

To etch the sample from the capping layer down to the top of the lower electrode layer (see Figure 3), a neutralized Ar ion beam of an ion beam etcher (IBE) is used. In a lab-assembled implementation of this tool, the process occurred over roughly 3 hours under high vacuum (base pressure 3e-8mbar). To detect the end etching point, an elemental analysis of the sample surface is done using Auger electron spectroscopy. After a typical 95-minute, five-step etching procedure with an etching dosage of about 94 mA.min of a Si|SiOx||Cr(5|Fe(50)|$C_{60}$(2ML)|CoPc(3ML)|$C_{60}$(5ML)|Fe(10)|Cr(100) stack (numbers in nm unless otherwise noted), the Auger scan shown in Figure 31 (a) was acquired. This high etching dosage did

not completely eliminate the 100 nm Cr capping layer. Three steps of further etching and Auger analysis (29 min, ~25 mA.min) shown in Figure 31 (b) reveal signals from one of the Fe electrodes and molecular spacer layer ($C_{60}$/CoPc). Ultimately, as seen by a higher Fe signal and a decreased C spectral intensity, a two-step etching (6 min, ~6 mA.min) in Figure 31 (c) reduced the molecular spacer layer to below its midpoint. The etching is then ended.

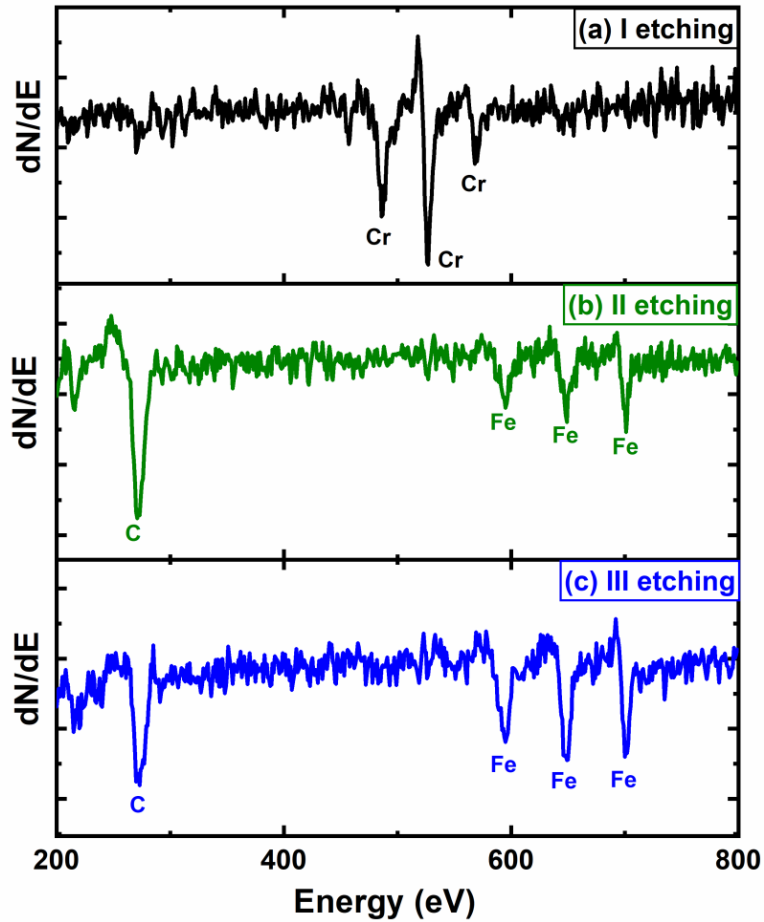

*Figure 31: Etching profile of multilayer stack Si|SiOx||Cr(5|Fe(50)|$C_{60}$(2ML)|CoPc(3ML)|$C_{60}$(5ML)|Fe(10)|Cr(100), after etching doses totaling (a) 94 mA.min, (b) 119 mA.min and (c) 125mA.min. The peaks at [272] eV, [489, 529, 571] eV, and [598, 651, 703] eV, and corresponds to C, Cr, and Fe respectively.[42]*

The sample is then transferred ex-situ to a high vacuum dielectric deposition chamber, in which 125nm $SiO_2$ is sputter deposited from a stoichiometric target in an Ar 25(sccm) /$O_2$ 5 (sccm) atmosphere with a radio frequency power of 85 watts and pressure is 4.2 mTorr.

### 6.3. Integrated Etcher/Dielectric growth system

To improve device processing efficiency, and eliminate the ex-situ transfer inherent to separate etching and encapsulation systems, we utilized an IBE tool at STNano that integrates endpoint detection using Secondary Ion Mass Spectrometry (SIMS) and in-situ dielectric sputter deposition capabilities. SIMS

Page **35** of 70

analyzes the mass of the atoms that are etched from the sample surface, so as to identify what is being etched. This real-time monitoring is not only time-efficient, but in principle guarantees accurate etching depth control.

Difficulties with SIMS detection can arise because Mask 2 may generate signal from parasitic elements that make it up (e.g. Fe in a stainless-steel mask). To avoid this issue, we recently turned to a Ti mask. Another difficulty is that our SIMS tool requires at least 7x7mm$^2$ in order to detect a signal. This issue is compounded by the per-element effective sensitivity. In our system, for instance, Fe exhibits a much weaker signal than Cr.

An example of this etching detection is shown in Figure 32. The etching conditions were as follows: Pressure = $2.9 \times 10^{-4}$ mbar, Forward RF Power = 98 W, Reflected Power = 1 W. At minute 18, we observe a decrease in the Cr intensity, while Fe begins to increase. This means that we have reached the top Fe/Cr interface. At 20 minutes etch time, we observe that the Fe signal starts falling, before increasing again. We conclude that the etching has reached the lower electrode. The etching is then stopped, and 125nm $SiO_2$ is sputter deposited in an Ar 25 (sccm) /$O_2$ 5(sccm) atmosphere with a radio frequency power of 85 watts and pressure is 4.2 mTorr. After that we can move to bead removal (see section2.4). As the etching and dielectric deposition procedures are carried out in the same system, the sample can be ultrasonicated in deionized (DI) water without the dielectric layer being compromised. As we will see, this switch from ex-situ to in-situ dielectric encapsulation is crucial because it results in bead removal with a success rate of 100% without damaging the dielectric.

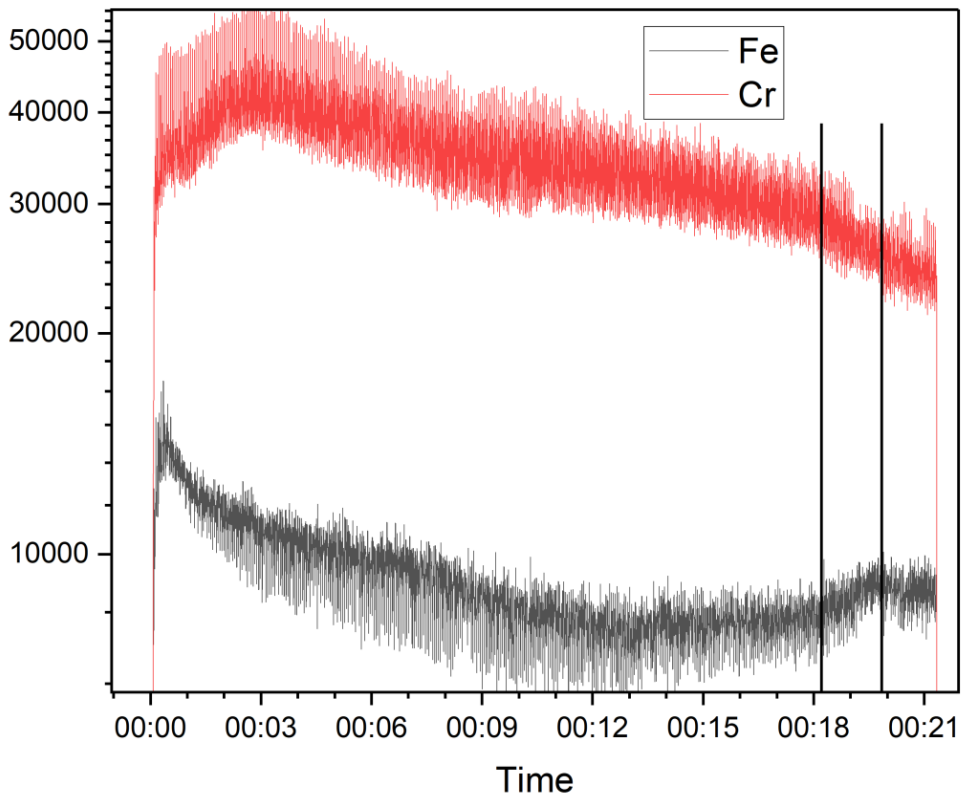

*Figure 32: SIMS Etching profile of a Si|SiOx||Cr(5|Fe(45)|C$_{60}$(1ML)|CoPc(4ML)|C$_{60}$(4ML/3ML)|Fe(15)|Cr(100) heterostructure.*

# 7. Nanobead Removal

A crucial yet tricky aspect of this process is removing the nanobeads after dielectric encapsulation (see Figure 5), and confirming that removal was successful. In this Section, we cover the various strategies that have been attempted and used.

### 7.1. The case of dry Dispersal

When dry dispersal of the nanobeads was used (see Section 5.1), a simple blowing with a nitrogen gun at a sharp angle of incidence with respect to the sample surface was used to remove beads. SEM was used to check for nanobead removal on test samples, but this was not done on actual device stacks to avoid charging issues.

## 7.2. The case of microfluidic deposition

Relative to the prior case of dry dispersal, removing nanobeads that have been deposited using microfluidic techniques appears to present additional challenges, though this could also reflect the additional work put into confirming removal in this case. In this subsection, we address the impact on removal effectiveness of 1) organic residue from the dried ink: strategies to minimize/clean, and the impact of SEM imaging; 2) functionalizing the nanobeads to control the chemical reaction with the sample surface; 3) various forms of mechanical energy, taking into account the perceived fragility of the sample against exposure to solvents/liquids. As we will see, a large parameter space of possible strategies, in combination, was tested. Surprisingly, at the time of this writing, it appears that in-situ dielectric encapsulation after etching (see Section 6.3) enables the use of ultrasonication in DI water and even acetone at 50C without degrading the $SiO_2$ layer. This is likely due to better $SiO_2$ adhesion onto the non-passivated etched surface, whereas ultrasonication using the two separate etching and encapsulation chambers (Section 6.2) would damage the dielectric coating.

Table 2 provides a summary of the testing space for process parameters and the various conditions evaluated for each parameter. We tested one parameter at a time, making reasonable attempts to explore all conditions thoroughly before moving on to the next parameter. During this process, we always kept the results of the previous parameters in mind to ensure that our findings were consistent and informed subsequent testing phases.

The first parameter we examined was the ink concentration (Parameter 1). As described earlier, we found that a 5:1 ethanol-glycerol mixture yielded the most effective results in terms of droplet formation (see sections 5.3.1.2 and 5.3.2.2). Given its superior performance, we decided to maintain this ink formulation as a constant in all subsequent experiments. Once this parameter was established, we proceeded to investigate the second parameter.

Droplet size (Parameter 2) plays a critical role in determining the process's ability to control the number of nanobeads per microdroplet, which in turn affects the formation of nanopillars bridging the macroelectrode junctions. We tested all four conditions for this parameter and our observations indicated that conditions 2 (droplet size 5-10 μm) and 3 (droplet size 10-15 μm) produced the most controlled and reproducible results. However, condition 1 (droplet size <5 μm) proved difficult to reproduce consistently, likely due to the instability of extremely small droplets. Condition 4 (droplet size >15 μm) resulted in an excessive bead count, which was undesirable for our intended application.

After finalizing the ink formulation and droplet size, we moved on to the third parameter: hot plate annealing conditions (Parameter 3). This parameter is crucial because it influences the presence and characteristics of organic residues left behind after the evaporation of microdroplets. These residues, if not properly managed, can act as unwanted macro-masks during the etching process, which could lead to larger sized pillars. To evaluate the effect of annealing, we first fixed the optimal ink formulation (Parameter 1) and droplet size (Parameter 2). We then tested nanobead removal using a nitrogen gun across different hot plate annealing conditions (Parameter 3). Our results demonstrated that bead removal was most effective when no annealing was performed (Condition 1). When a sample is subjected to annealing, the elevated temperatures cause the evaporation of any remaining volatile compounds, which can result in increased surface adhesion. This occurs due to thermally induced changes such as oxidation, polymerization, or other surface reactions that strengthen the bond between the droplet and the substrate. Additionally, annealing can lead to localized restructuring of the substrate surface, potentially creating stronger intermolecular or electrostatic forces that further anchor the beads in place. As the temperature and duration of annealing increase, these effects become more pronounced, making bead removal increasingly difficult. To validate our findings, we performed extensive testing and exploring numerous samples under different combinations of temperature and time for annealing. Despite multiple attempts using a nitrogen air gun for bead removal after each parameter, the best success rate we achieved was less than 2%, as shown in Figure 33 .

To increase the bead removal success rate, we moved to the next parameter which is the $O_2$ plasma cleaning duration (Parameter 4). To determine if prolonged plasma cleaning enhances bead removal by improving surface cleanliness and reducing organic residues by keeping earlier conditions in mind. Longer plasma cleaning durations facilitated easier bead removal, likely due to better surface cleaning as shown in Figure 36. However, despite testing multiple samples without ultrasounds, the success rate did not exceed 10% as shown in Figure 34. We observed in later testing that waiting 12 hours in ambient conditions leads to a better reduction in the area of organic residue, compared with loading the sample into the IBE chamber immediately after nanobead deposition.

In subsequent testing, we used optical inspection after IBE/encapsulation as an indirect means to determine whether organic residue and accompanying nanobeads have been removed. Limited testing suggests that prolonged acetone baths with US and acetone heating to 50C can improve this metric.

We then proceeded with subsequent device fabrication steps, with working junction statistics detailed in section 10. To further increase the success rate, we moved to the next parameter: the time between bead positioning and dielectric deposition (Parameter 5) and tested the role that a time delay between bead deposition and dielectric deposition plays (e.g. because bead adhesion changes over time). We observe no significant improvement in bead removal success, indicating that dielectric deposition can proceed (see section 2.3) within a week of bead printing without affecting results.

The next parameter was the thickness of dielectric (Parameter 6), to evaluate if thinner dielectrics ease bead removal while maintaining leakage current performance. We tried several conditions (1-3) 135nm-80 nm and concluded that thinner dielectrics (closer to 80 nm) showed marginally easier bead removal but risked higher leakage currents. Despite optimization, the success rate remained near 10%, confirming that dielectric thickness alone was not the limiting factor.

Our next parameter SEM (Parameter 7) which is imaging technique to determine if electron beam exposure during SEM imaging alters bead adhesion. We observed that bead removal was slightly harder after SEM imaging, suggesting that electron beam exposure may strengthen bead adhesion. However, the maximum success rate remained around 10% as shown Figure 35.

To increase the success ratio, we tried the next parameter: swirling in ethanol with annealing (Parameter 8) to test if solvent exposure combined with thermal treatment loosens bead adhesion. We varied ethanol swirling durations and annealing temperatures (Conditions 1-8) with no measurable improvement in bead removal success. After that we tried ultrasonication in DI water (Parameter 9) for different durations (Conditions 1-3) to exploit mechanical agitation for bead dislodgment. We observed that beads were removed, but that the dielectric layer was damaged, rendering this method unsuitable as shown in Figure 37. Note that this test was performed on samples with ex-situ dielectric encapsulation.

We tested the next parameter, which is ultrasonication using our wire bonder (Parameter 10) to apply controlled mechanical force for bead removal. We tried gradual increase in force and rounds (Conditions 1-6). Beads were successfully removed, but the dielectric sustained irreparable damage, similar to (Parameter 9).

The next parameter is blowing beads using a nitrogen gun (Parameter 11), i.e. a high-pressure gas for non-contact bead removal and varied the duration (Conditions 1-5). We found that longer blowing durations improved success marginally, but sample handling remained critical. This method was applied after testing every other parameter (pre-SEM imaging) as a baseline check. As the bead removal success ratio is still around 10 % we tried lens paper (Parameter 12) to mechanically dislodge beads via gentle abrasion. Here, bead removal was inconsistent, and the method often scratched the sample as shown in Figure 38. Reproducibility was poor, making it impractical.

The next parameter is Monodisperse carboxyl functionalized beads 5% w/v ethanolic suspension (Parameter 12). The strategy was to chemically weaken bead adhesion via pH treatment and Immersion in pH 10 solution for 12 and 24 hours. We obtained 100% bead removal success, but the sample was severely damaged as shown in Figure 39, likely due to prolonged chemical exposure to the alkaline pH 10 solution, which induced structural degradation of the sample material.

To systematically optimize bead removal, each parameter was tested under multiple conditions, with bead removal efficiency immediately verified using a nitrogen gun followed by SEM imaging to confirm complete bead detachment. This rigorous validation process required preparing and testing numerous identical samples, making it particularly challenging due to the time-consuming nature of SEM analysis and the need for precise comparative assessments. After each parameter adjustment, the nitrogen gun test provided initial removal feedback, while SEM imaging delivered definitive confirmation of success or failure. This iterative approach, though labor-intensive, was crucial for isolating key variables, as some methods (like ultrasonication) removed beads but damaged the dielectric, while others (like extended plasma cleaning) showed only marginal improvements. Ultimately, this exhaustive testing protocol led to the identification of the optimal solution: in-situ pillar encapsulation see section 6.3 followed by DI water ultrasonication, potentially augmented by prolonged (12-36 hours) acetone bath including sequences at 50C. This helped achieve a much larger success at bead removal without compromising dielectric integrity.

|  | Parameters | Condition 1 | Condition 2 | Condition 3 | Condition 4 | Condition 5 | Condition 6 | Condition 7 | Condition 8 |
|---|---|---|---|---|---|---|---|---|---|
| 1. | Ink Concentration | Ethanol-Glycerol (5:1) | Ethanol-Glycerol (7:1) | Ethanol-Glycerol (10:1) | Ethanol-Glycerol (5:2) | | | | |
| 2. | Size of droplet | <5 µm | 5-10 µm | 10-15 µm | >15 µm | | | | |
| 3. | Hot Plate Annealing Conditions (Temp & Time) | Without annealing | 60°C, 10-60 Min (steps) | 70°C, 10-60 Min (steps) | 80°C, 10-60 Min (steps) | 90°C, 10-60 Min (steps) | 100°C, 10-60 Min (steps) | | |
| 4. | $O_2$ plasma cleaning duration | 10 Min | 20 Min | 30 Min | 40 Min | 50 Min | 1 Hour | 1-20 Hours (steps of 1 hr) | |
| 5. | Time between bead positioning and dielectric deposition | Same day | 1 day | 4 days | 1 week | | | | |
| 6. | Dielectric thickness | 135 nm | 110 nm | 80 nm | | | | | |
| 7. | SEM secondary vacuum without providing energy | EHT on | EHT off | | | | | | |
| 8. | Swirling in Ethanol with annealing | 40°C (10-120 Sec, step 10) | 60°C (10-120 Sec) | 80°C (10-120 Sec) | 100°C (10-120 Sec) | 120°C (10-120 Sec) | 140°C (10-120 Sec) | 180°C (10-120 Sec) | 200°C (10-120 Sec) |
| 9. | Ultrasonication in DI water | 1 sec | 2 sec | 5 sec | | | | | |
| 10. | Ultrasonication of wire bonder (# of rounds) | 5 rounds | 10 rounds | 12 rounds | 15 rounds | 20 rounds | 25 rounds | | |
| 11. | Nitrogen Gun | 5 sec | 10 sec | 15 sec | 30 sec | 60 sec | | | |

| 12. | Lens Paper | 1 Rub | 2 Rub | 3 Rub | 4 Rub | 5 Rub | | | |
|---|---|---|---|---|---|---|---|---|---|
| 13. | Carboxyl functionalized beads | Base 10 solution for 24 hours. | Base 10 solution for 12 hours. | | | | | | |

*Table 2: Parameters and Conditions for bead removal without ultrasounds.*

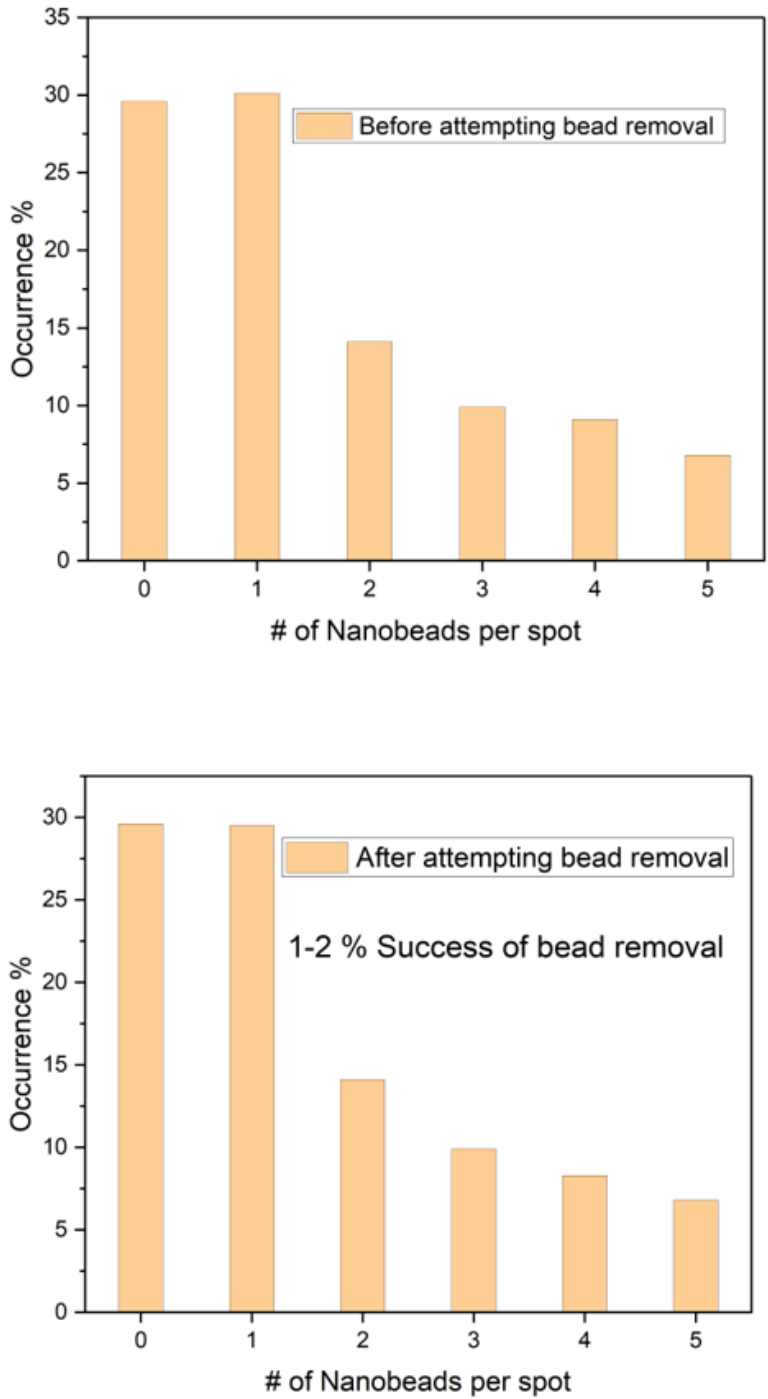

*Figure 33: Statistics of bead removal success before and after trying (1-3) parameters. Parameter 11 nitrogen is used every time before testing the bead removal.*

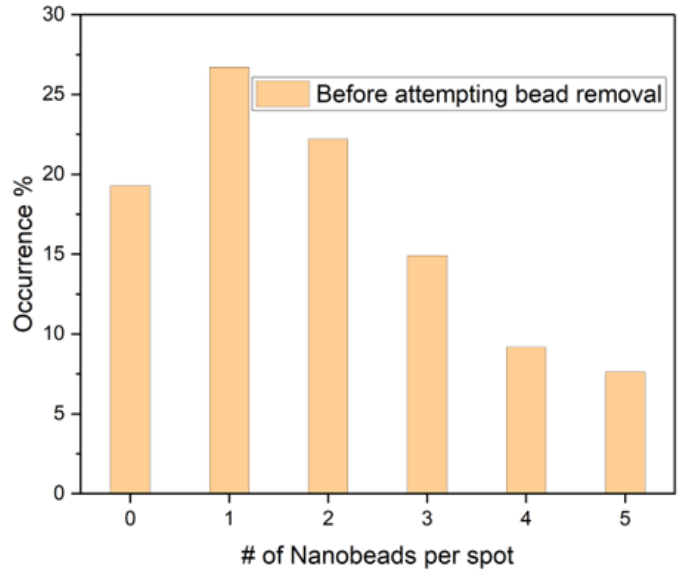

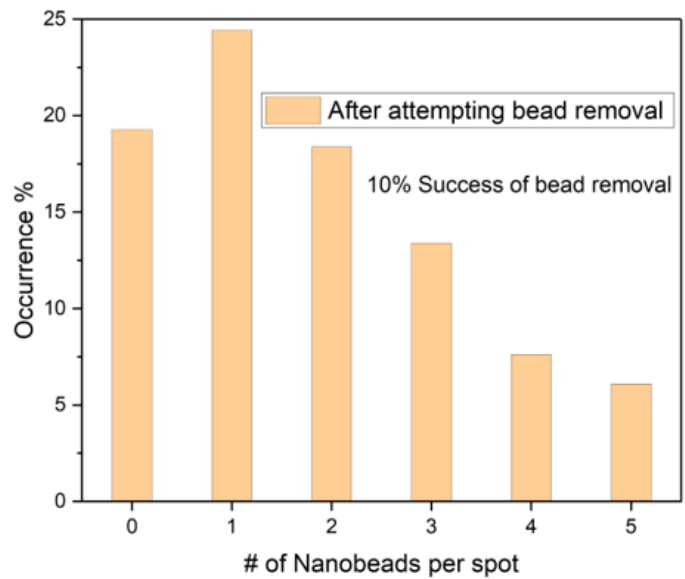

*Figure 34: Statistics of bead removal success before and after trying (1-4) parameters with increased success percentage of 10%. Parameter 11 (nitrogen gun) is used every time before testing the bead removal.*

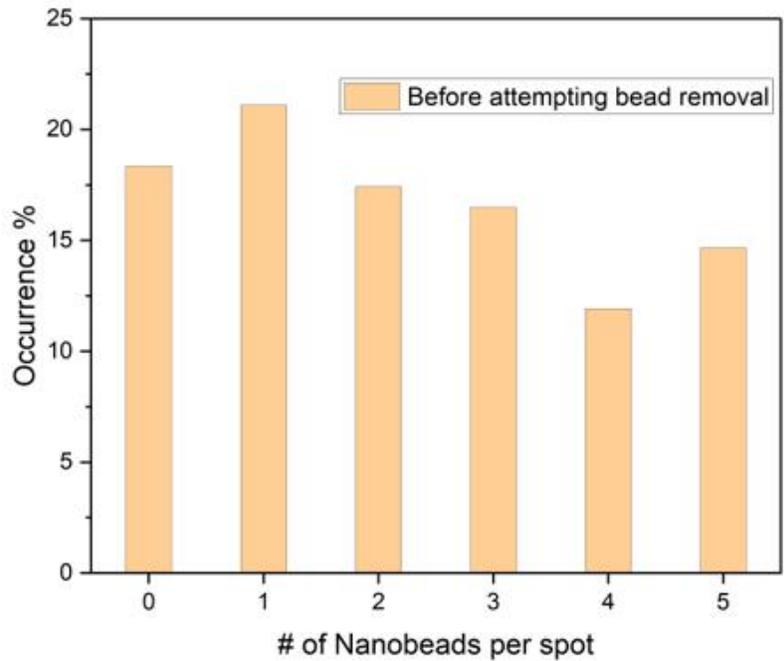

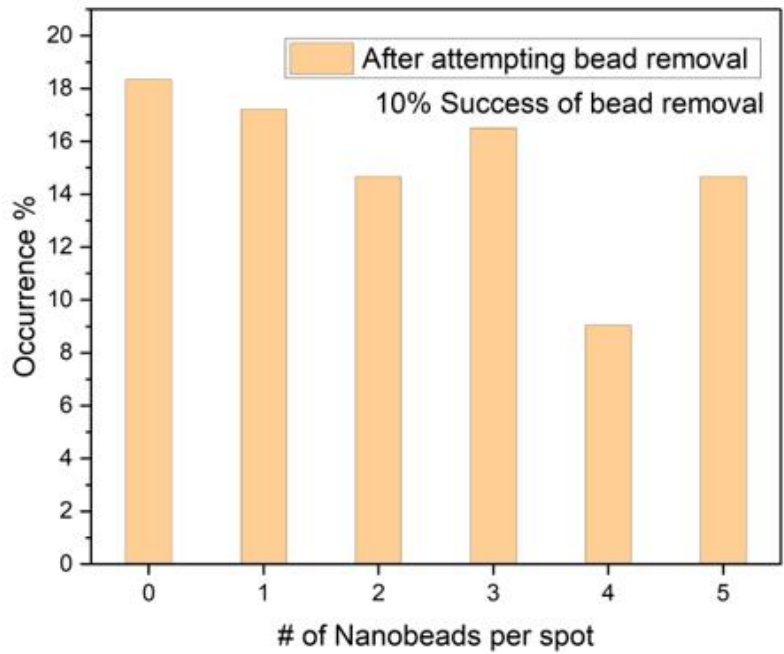

*Figure 35: Statistics of bead removal success before and after trying (1-7) parameters. The success rate is still 10%. Parameter 11 (nitrogen gun) is used every time before testing the bead removal.*

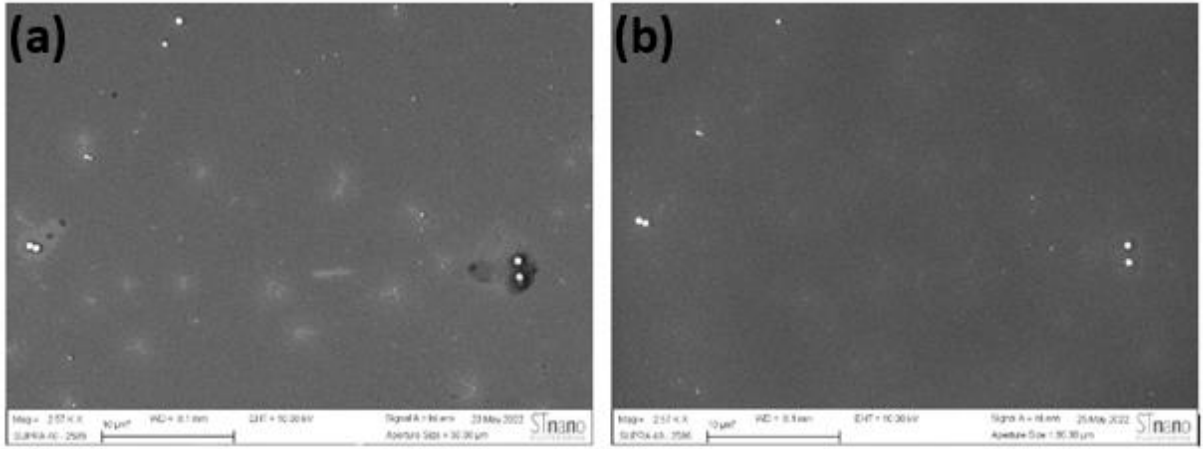

*Figure 36: SEM images of a test sample (a) before and (b) after O$_2$ plasma cleaning (Parameter 4).*

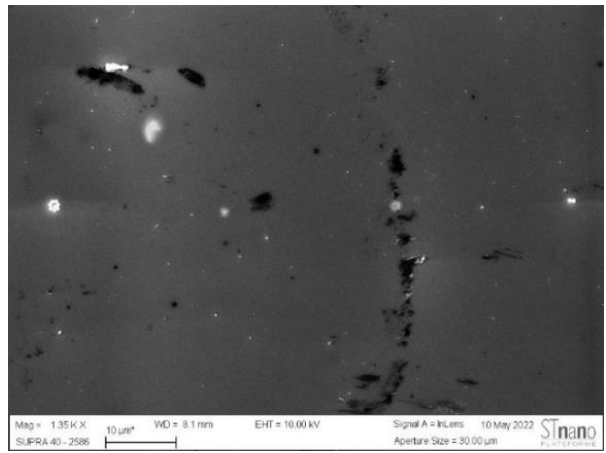

*Figure 37: SEM image of a test sample after ultrasonication using the wire bonder (Parameter 10).*

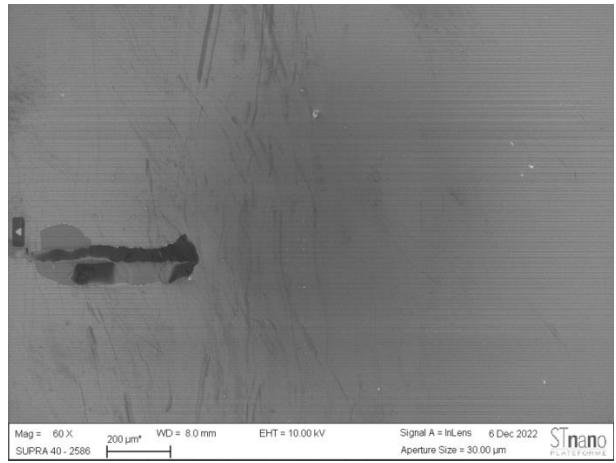

*Figure 38: SEM image after Lens paper (Parameter 11) was used in an attempt to remove beads. Its use led to surface deterioration.*

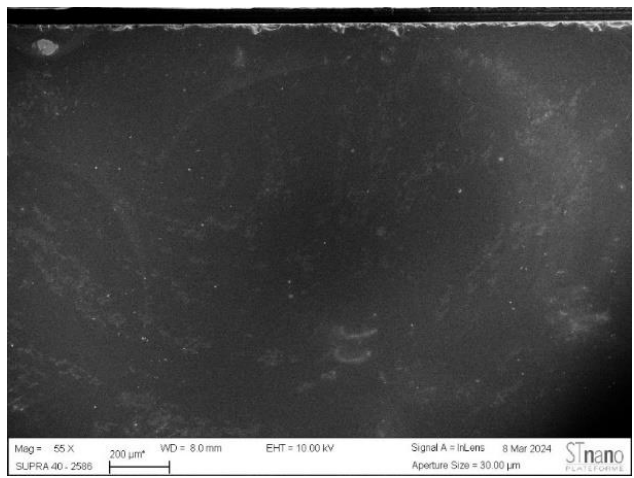

*Figure 39: SEM image after Carboxyl-functionalized beads (Parameter 12) were treated in pH 10 solution to weaken adhesion. Although bead removal was successful, the sample was damaged due to prolonged alkaline exposure.*

# 8. Technological top Electrode

Once etching of the nanobead-decorated sample, dielectric encapsulation and nanobead removal are completed, a technological top electrode is deposited to enable electrical access to every junction's nanopillar(s) (see Figure 6). Given the junction pitch (i.e. the distance between junctions) and the process's design geometry, a good alignment is required in order 1) to obtain top electrodes that intersect the lower electrode at 2) the predetermined location. This second point is crucial if the nanopillars are explicitly positioned against a nominal position dictated by the overlap of the process's masks. This in turn hinges on the maximum x-y offset in reproducibly positioning each succeeding mask.

## 8.1. Loose Mask 3 Implementation

If Mask 3 is hand aligned, this is time consuming and more prone to errors as mentioned before for mask 1 and mask 2 (see section 3.2.1 and 6.1.1). An example of a well-executed top electrode metallization is shown in Figure 40.

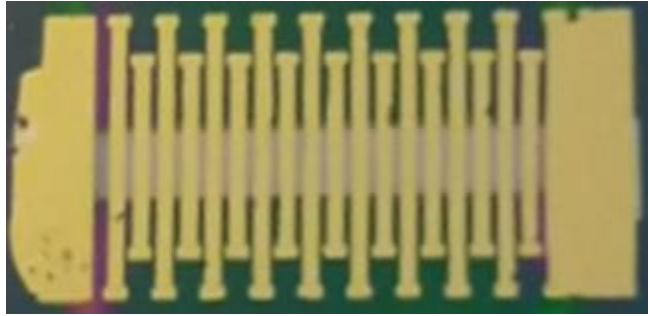

*Figure 40: Optical image of a cell upon process completion using the loose mask implementation.*

## 8.2. Auto-aligned Mask 3 Implementation

In the auto-aligned implementation, Mask 3 as shown in Figure 41 is used to define the top electrode metallization contacts. This implementation assumes an error in x-y repeatability that is at most 40 microns. Furthermore, a nanobead is positioned within 5 microns of the ideal location because of a 10-micron droplet size. These considerations mean, to ensure that the process shall work, the top electrode width should be at least twice these systematic alignment errors. We thus chose a width of 85 microns.

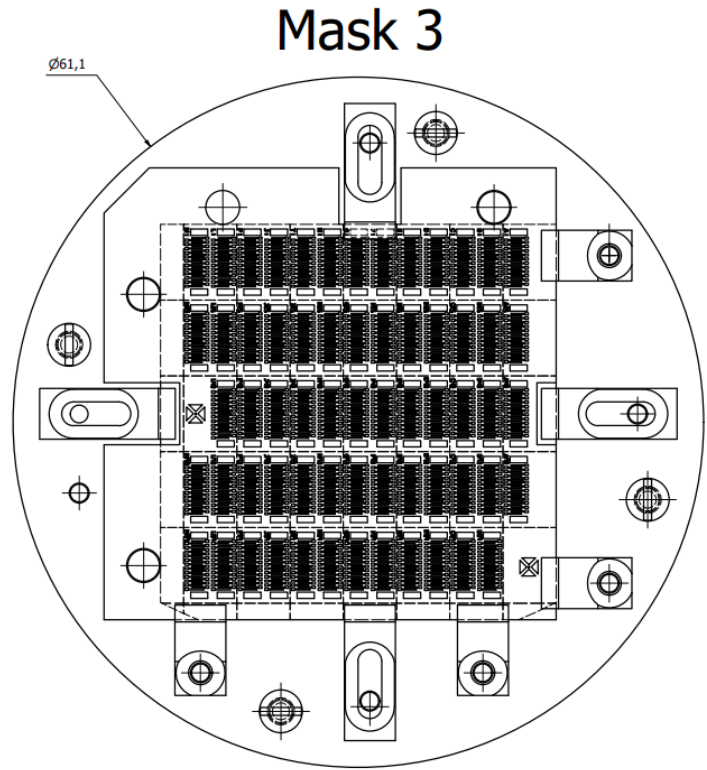

*Figure 41: Auto-aligned process: schematic of sample plate, sample and Mask 3 before top electrode metallization (see section 2.5) Multiple clamps are used to reduce the shadowing effect.*

As with Mask 1, the use of additional clamps and/or Kapton tape is desirable in order to obtain correct masking. An example of the as-prepared sample with auto-aligned Mask 3 prior to top electrode deposition is shown in Figure 43, and a zoom on one completed cell is shown in Figure 43.

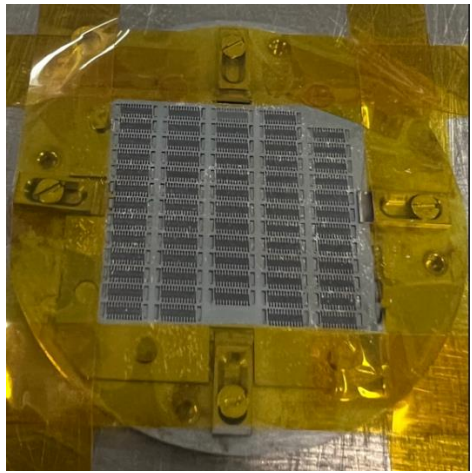

*Figure 42: Auto aligned process: Image of sample plate, sample and mask 3 on top for metallization. The picture was taken before top electrode deposition. Kapton tape was used to reduce the risk of mask buckling, and also to prevent*

*additional material from being deposited onto the sample holder and clamps, which can influence SIMS detection in a subsequent process.*

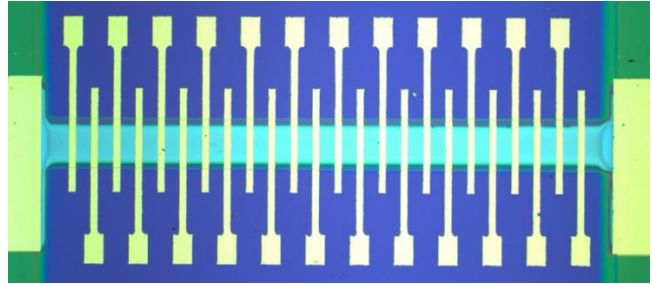

*Figure 43: Optical image of a cell after step 5 (see section 2.5) Auto-aligned mask 3 implementation.*

After step 5 (see Section 2.5), the implementation of auto-aligned mask 3 significantly improved masking accuracy, resulting in reduced misalignment, minimized shadowing, and uniform junction pad thickness. This result is tempered by a switch in mask supplier, which introduces inaccuracies and thus mask buckling, leading to a relative loss of definition. However, this didn't cause top electrodes between adjacent junction to overlap.

# 9. Sample Dicing

In the loose mask approach described in Sections 3.2.1, 6.1.1 and 8.1, care is taken to implement the 5-step technological process on Si wafers that are already precut to the maximum size possible for our magneto transport setup. In fact, that size, 5x7mm, conditions the actual lower electrode and top electrode design, with wire-bonding induced constraints on the 100-micron width of the top electrode contacting pad.

Working with 35x35mm wafers per the auto-aligned approach described in Sections 3.2.2, 6.1.2 and 8.2 requires a strategy to dice the final processed wafer. Dicing normally requires that the sample be coated with resist to protect the metallized top electrodes from being scratched by dicing debris. As described in Section 7.2, our initial testing on nanobead removal after dielectric encapsulation using a separate tool (see Section 6.2) indicated that solvent baths and ultrasonication would cause the encapsulation layer to rapidly break off. As we later realized, this was likely due to surface passivation issues in ambient conditions between the etching and encapsulation steps when separate tools are used. Indeed, later testing using the integrated Etcher/encapsulation tool (see Section 6.3), performed to improve the nanobead removal success rate (see Section 7.2) revealed that only minor damage to the dielectric layer occurs upon ultrasonication. This enabled the possibility to dice the sample in normal fashion.

However, prior to this technical breakthrough, we performed an extensive investigation into a technological strategy that involves deep reactive ion etching, into the Si substrate, of grooves (see Section 9.1. These grooves are intended to facilitate the cleaving of the sample through mechanical means, using a custom-designed tool (see Section 9.2). To be clear, the remainder of Section 9 is as of this writing technologically deprecated, but we include it for the sake of completeness of our research into this vertical nanojunction molecular technological process.

## 9.1. Substrate precut process

To circumvent the standard resist+dicing procedure, we implemented a pre-cut process on our virgin Si/SiOx// 35x35 mm wafer (see Figure 44). First, we deposited 100nm of chromium in our sputtering chamber via magnetron sputtering on a silicon substrate (35x35mm). We then applied a thin layer of AZ5214 resist onto the substrate using a spin coater (4000rpm/30s) + hot plate anneal at 100C for 60s. UV Laser lithography and a developer solution (AZ 726MIF for 60s) was used to create grooves in the resist layer (panel b). The chromium layer was then wet-etched, and $SiO_2$/Si were etched using reactive Ion Etching (see panels c-d). After the removal of resist and chromium, the substrate is left with 85 µm deep pre-cuts as schematized in panel (e). Only then does the process begin with Step 1 (see Section 2). An image of the completed pre-cut substrate is shown in Figure 45, and a depth profile, acquired using profilometer, of the precut is shown in Figure 46.

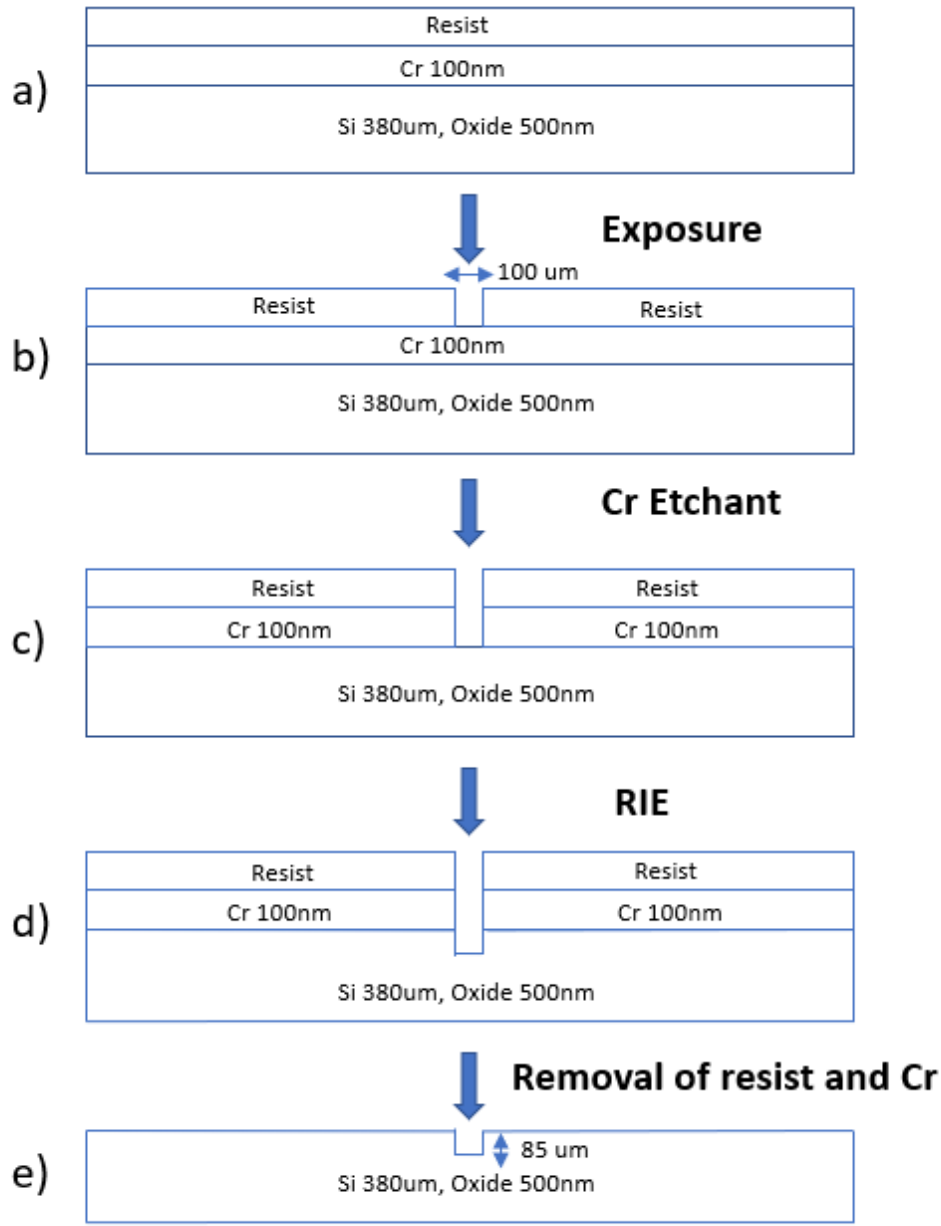

*Figure 44: Pre-cuts Process description on our silicon substrate: (a) Silicon substrate with Cr deposited and Resist. (b) after patterning with UV lithography. (c) After removing unprotected Cr with etchant. (d)Etching silicon oxide and silicon substrate with RIE. (e) After removal of resist and Cr, silicon substrate with precut.*

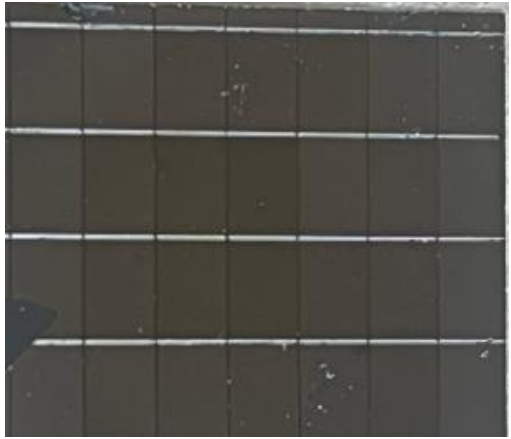

*Figure 45: Optical image of our 35x35 mm substrate with Pre-cut depth of 90um, each cell is 5x7mm to fit on our FERT chip, image of substrate with pre-cut before step 1. See section 2.1*

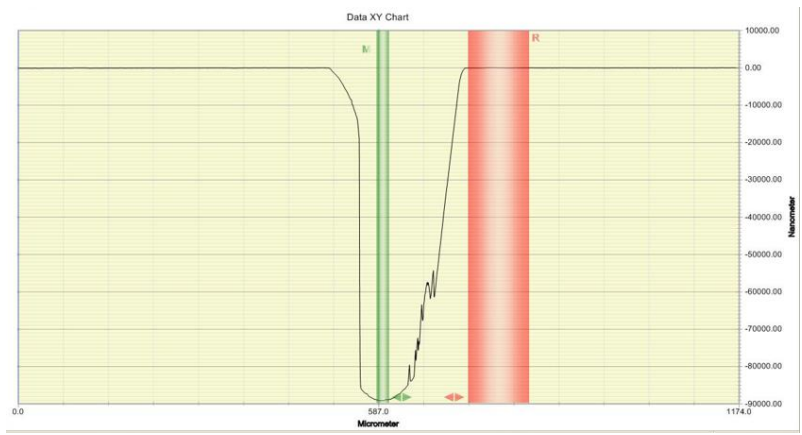

*Figure 46: Precut depth of 90 μm measured using a profilometer before step 1 . See Section 2.1.*

## 9.2. Custom substrate cleaver

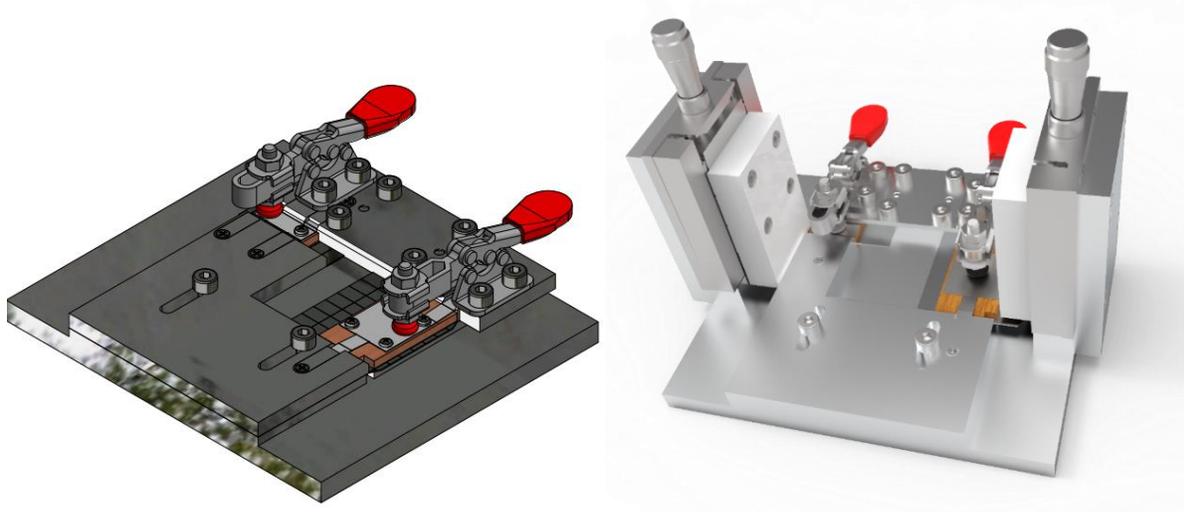

*Figure 47: 3D images of a custom-designed and built cleaving tool for precise sample segmentation.*

Once the 5-step process is completed on a sample with pre-cuts of the substrate, the sample must then be cleaved along these grooves. To do so, we designed and assembled a custom tool (see Figure 47). Its design enables the square substrate to first be cut into rows (right-hand area). Then, each strip is inserted into the left-hand holder in order to dice every row into the required approx. 5x7 samples for use on our custom chips in our magnetotransport setups. To hold the sample in place with minimum scratching or electrostatic discharge on the top surface, wood was used to apply even pressure on the sample. The wood pieces are glued to thin metal sheets attached on either side in order to automatically raise the wood pieces when not pressed down. The sample portion to be cleaved hangs over a well-defined edge (either left or right). The cleaver consists in a polyoxymethylene block mounted on a micromanipulator. It is positioned 1mm from this edge. This allows the precut to be positioned immediately over the edge of the sample holder surface, and the cleaving edge on the other side of the precut.

Repeated testing of this tool on 380-micron thick Si wafers led to incremental improvements on sample handling, holding with spring-loaded wood pieces, additional grooves in order to grab the sample from under the wood pieces. However, our repeated testing revealed that, at this stage of process development, we were unable to reproducibly dice the completed junction substrate into the required 5x7mm pieces. While the substrate would dice along the precut most of the time, catastrophic fractures outside the precut lines could also occur. We presume, without proof, that unless a thinner substrate were used, the precuts would have to be deeper (i.e. a greater ratio of the nominal Si substrate thickness) in order to eliminate this shortcoming. To do so would require substantial modifications to the chromium and resist layer thicknesses, since the present thicknesses were just sufficient to achieve the 85 micron-deep grooves.

# 10. Prober Results

Achieving a high success rate of working junctions with a low distribution of resistance is the ultimate goal of this process. This relies first on a high-quality stack and process. Here are some factors that contribute to the effective success rate are:

1) Deposition of the heterostructure's metallic counter electrode onto the sample while at room temperature. As discussed elsewhere[53], this can introduce an interdiffusion of metal atomic species that can reduce the junction resistance or even short-circuit it. We discuss buffer-layer assisted growth[54,55] in our Perspectives (see Section 11).
2) The cleanliness of the sample a) prior to and b) after heterostructure deposition (Mask 1); and c) prior to and d) after top electrode metallization (Mask 3). This is in large part due to residues on Masks 1/3 from deposition on prior samples. At the time of this writing, we have confirmed that our masks can withstand an ultrasonic bath. This is useful because these residues can impair heterostructure homogeneity through shadowing, create unwanted masking in future steps (e.g. unwanted pillar structures upon IBE), or even generate catastrophic electrical bridges between the junction's macro top and bottom electrodes.
3) The presence of organic residue from the ink droplet during printing (see Section 2.2) can adversely affect the effective junction area. To mitigate this issue, annealing should be avoided as previously discussed (see Section 7.2). Additionally, oxygen plasma cleaning prior to Step 3 (see Section 2.3) can further minimize contamination. Limited testing shows that the organic cap can be removed thanks to hot acetone ultrasonic baths after dielectric encapsulation.
4) The nanobead count distribution profile. Our microdroplet printing is typically tuned to yield almost half of the spots with precisely 1 nanobead, but this causes 1/3 or more spots to not have a nanobead. In this case, if any remaining organic residue doesn't conduct (well at least) and/or is encapsulated by $SiO_2$, then the resulting junction will be open-circuit.
5) Nanobead removal appears to work, but has been tested at Steps 2 (see Sections 2.2 & 5) and 3 (but prior to dielectric encapsulation; see Sections 2.3 & 6), and cannot systematically be tested upon Step 4 (see Section 2.4 & 7) because SEM imaging is next to impossible on an insulating sample.
6) The ability to have positioned nanobeads, and thus defined pillars, within the areal overlap between the junction's macro top and bottom electrodes. This point is especially prevalent in the dry dispersal approach. However, we point out that the following results using the auto-aligned mask approach were performed using a random linear array of microdroplets (see Section 5.3.2.3), with a pitch at least ½ smaller than the top metallization electrode pitch, to ensure that the nanobead(s) of at least one microdroplet would land within the macro junction area. At the time of this writing, limited testing suggests that a one-droplet-one-junction approach can yield results.

A prober instrument was utilized to examine the resistances at 5-10mV applied bias voltage, and sometimes the IV characteristics of our devices. Two-point (2-pp) or four-point (4-pp) probe methods are among the most widely used methods for measuring electrical resistance. The two tungsten carbide needles function as voltage and current measuring probes in a 2-pp mode. The measurement loops in this case always account for the bottom electrode resistance contribution, which has the potential to alter the total resistance resulting from the individual junction electrodes. For accurate measurements of low-resistance junctions (R<1kOhm), 4-pp mode is helpful. In a 4-pp, two probes have a known bias voltage put between them, and the other two probes are used to measure the current that results. We generally used the 2pp technique, noting that our lower electrode is expected to have R≈100Ohm and our top electrode R≈10 Ohm. In what follows, only those devices with a junction resistance at least 2x greater than the bottom electrode have been considered to be working.

The statistic of junction resistance is shown in Figure 48 for a Si/SiO$_x$/Cr(5nm)/Fe(50nm)/C$_{60}$(1ML)/CoPc(3ML)/C$_{60}$(5ML)/Fe(10nm)/Cr(100nm) stack. The estimated number of beads per junction is less than 3, the bead diameter is 500nm, and a 30S tip was used to print droplets with a size of 8-10 um. Etching was done until the barrier, followed by in-situ dielectric deposition of 125 nm of SiO$_2$ . The bead removal is done with the help of ultrasonication in DI water for 1-2 minutes.

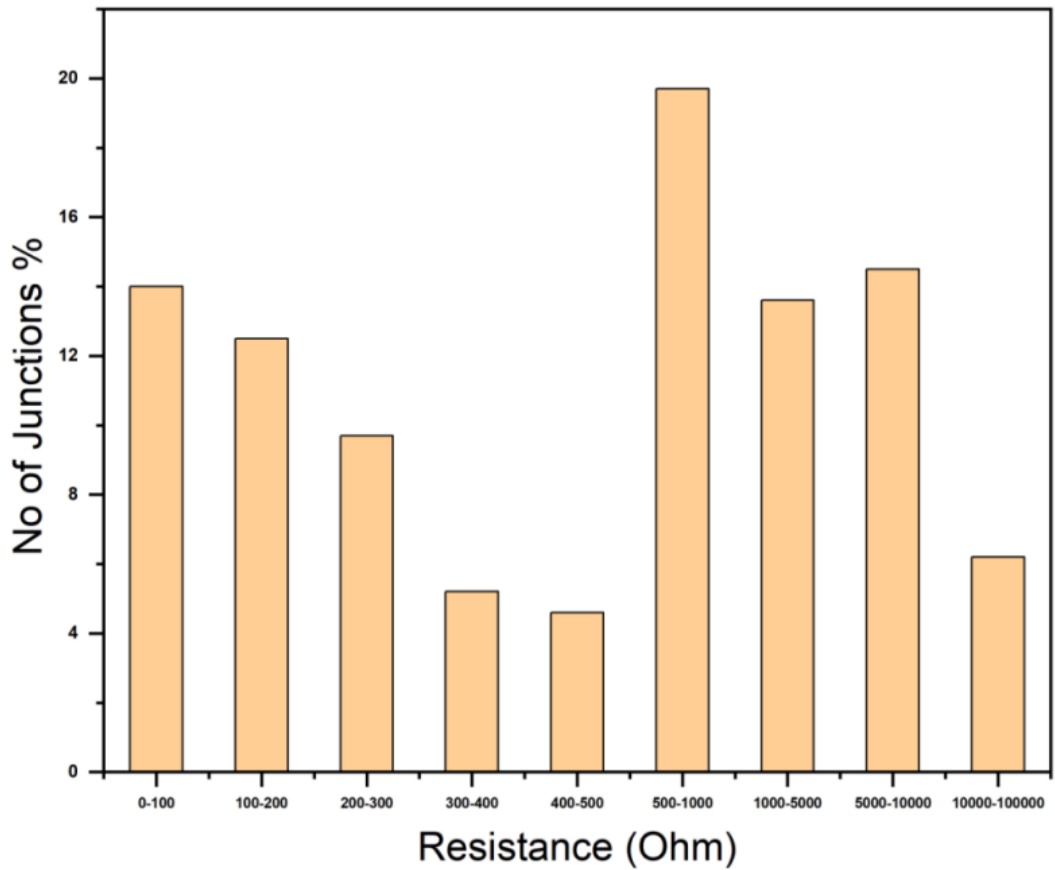

*Figure 48: Resistance statistics for a Si/SiO$_x$/Cr(5nm)/Fe(50nm)/C$_{60}$(1ML)/CoPc(3ML)/C$_{60}$(5ML)/Fe(10nm)/Cr(100nm) stack*

## 10.1. First microfluidic positioning

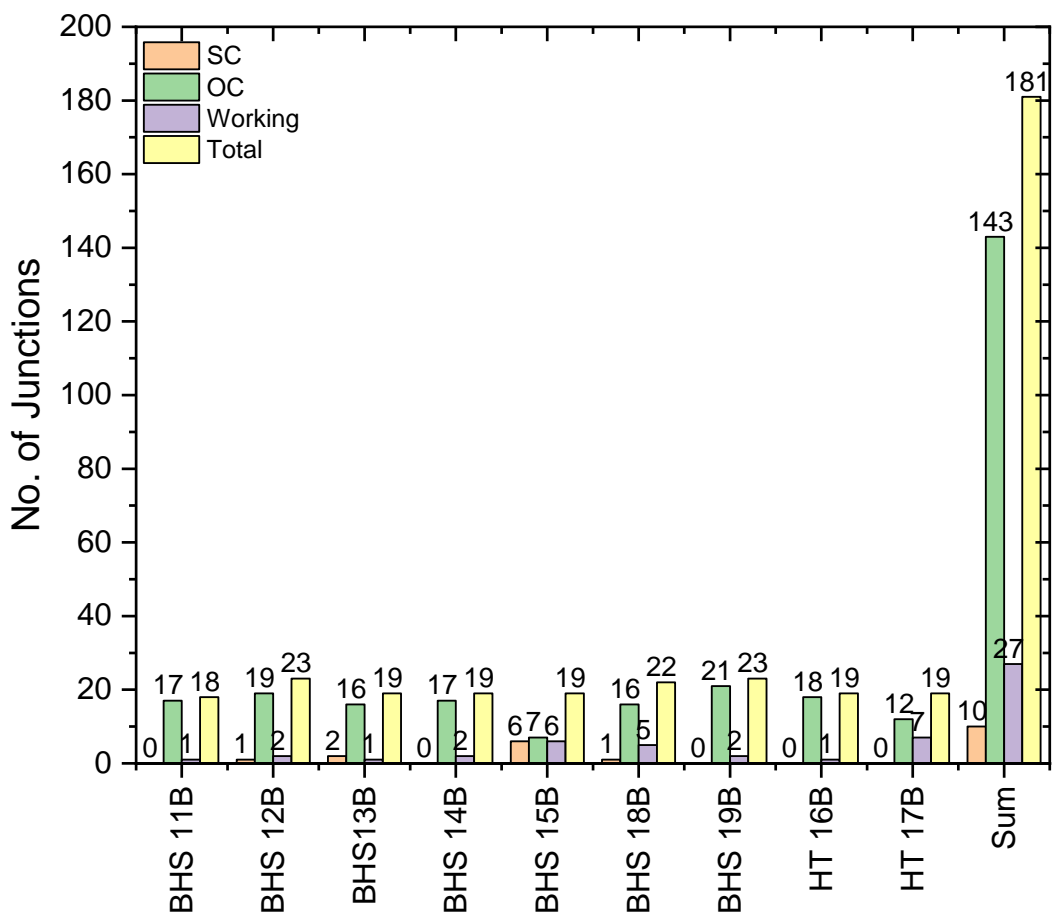

*Figure 49: Devices statistics using the 'microfluidic pen lithography' technique. Graphical representation of the prober results for the nanopillars fabricated from a series of capped hybrid trilayer stacks. The X-axis represents the beads processed capped hybrid sample, while the Y-axis shows the frequency of short-circuited (pink), open-circuited (green), working (purple) and total (yellow) nanopillars fabricated from different capped trilayer hybrid stacks. SC*

*corresponds to Junction resistance < top/bottom metal electrode, i.e. <100 Ohm, OC corresponds to Junction Resistance = GigaOhm/floating higher resistance values (due to the relative imprecision of the Keithley 2400), Working Devices = kilo Ohm-mega Ohm.[42]*

We performed the nanobead process using the initial microfluidic pen lithography technique (see Section 5.3.1) on a series of capped hybrid trilayer stacks and found a good success rate of ~15%. All these electrical or probing measurements were performed in 4-point contact mode on the prober machine. Before the prober measurements, the Keithley 2400 sourcemeter was also calibrated with the standard resistances. Figure 49, provides a graphical presentation of the prober data for the capped hybrid nanopillars fabricated from a series of capped hybrid trilayer stacks. As a note, the K2400 measures resistances reliably only up to ~50MOhm.

## 10.2. Auto-aligned mask process

Bead removal for devices TZ240823 and TZ031023 was performed using a nitrogen gun without ultrasonication, whereas for device TZ190224, bead removal was conducted via ultrasonication in DI water (see Section 7.2). We observed that the percentage of working junctions in device TZ190224 is higher (26%) compared to TZ240823 (4%) and TZ031023 (18%). The total number of junctions in TZ240823 is lower than 1512 due to technical difficulties encountered at the initial stages of fabrication (see Section 3.2.2 for details). In contrast, for device TZ190224, the sample was not entirely printed, resulting in a total of 1176 junctions. The statistical analysis of working junctions is presented in Figure 50.

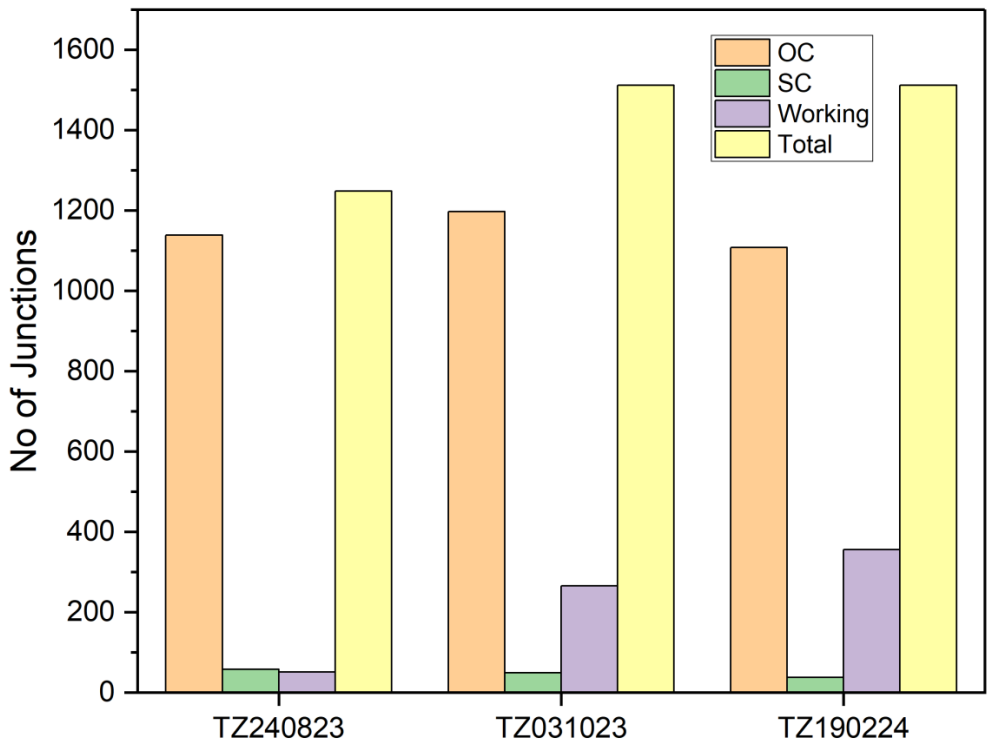

*Figure 50: Statistics of junctions using the microfluidic process. The same stack composition was used for TZ240823, TZ031023 & TZ 190224 samples: Si/SiO$_x$/Cr(5nm)/Fe(50nm)/C$_{60}$(1ML)/CoPc(3ML)/C$_{60}$(5ML)/Fe(10nm)/Cr(100nm) The frequency of short-circuited (pink), open-circuited (green), working (purple) and total (yellow) nanopillars fabricated from different capped trilayer hybrid stacks is shown. Bead removal for sample TZ 240823 & TZ 031023 were done with the help of nitrogen gun whereas bead removal for sample TZ 190224 were done with the help of ultrasonication in DI water for 1-2 minutes.*

To evaluate the effects of aging, resistance measurements were repeated after a period of 3–4 months to observe any changes over time. In most cases, the resistance values remained relatively stable, indicating good long-term stability. However, in some instances, noticeable variations were observed, with resistance either increasing or decreasing. Examples of such junctions with changes are shown below

| Junction | Prober March 2024 | Prober July 2024 |
|---|---|---|
| A29 | 56K | 47K |
| A44 | 73K | 80K |
| B21 | 305 | 428 |
| B23 | 270 | 508 |
| B103 | 331 | 1K |
| B107 | 232 | 280 |
| B109 | 1K | 1.3K |
| B111 | 811 | 530 |
| B113 | 17K | 1.6K |
| B117 | 654 | 6K |
| B119 | 2K | 1K |
| B118 | 2.6K | 3K |
| B114 | 45K | 43K |
| B112 | 355 | 2K |
| B110 | 237 | 287 |
| B108 | 213 | 300 |
| B106 | 747 | 980 |
| B104 | 1.3 | 2K |
| B102 | 341 | 1K |

Table 3. We have observed that up to a 10% change in device resistance can result from differing contacts.

| Junction | Prober March 2024 | Prober July 2024 |
|---|---|---|
| A29 | 56K | 47K |
| A44 | 73K | 80K |
| B21 | 305 | 428 |
| B23 | 270 | 508 |
| B103 | 331 | 1K |
| B107 | 232 | 280 |
| B109 | 1K | 1.3K |
| B111 | 811 | 530 |
| B113 | 17K | 1.6K |
| B117 | 654 | 6K |
| B119 | 2K | 1K |
| B118 | 2.6K | 3K |
| B114 | 45K | 43K |
| B112 | 355 | 2K |
| B110 | 237 | 287 |
| B108 | 213 | 300 |
| B106 | 747 | 980 |
| B104 | 1.3 | 2K |
| B102 | 341 | 1K |

*Table 3:Comparison of resistances measured with prober in March 2024 and July 2025.*

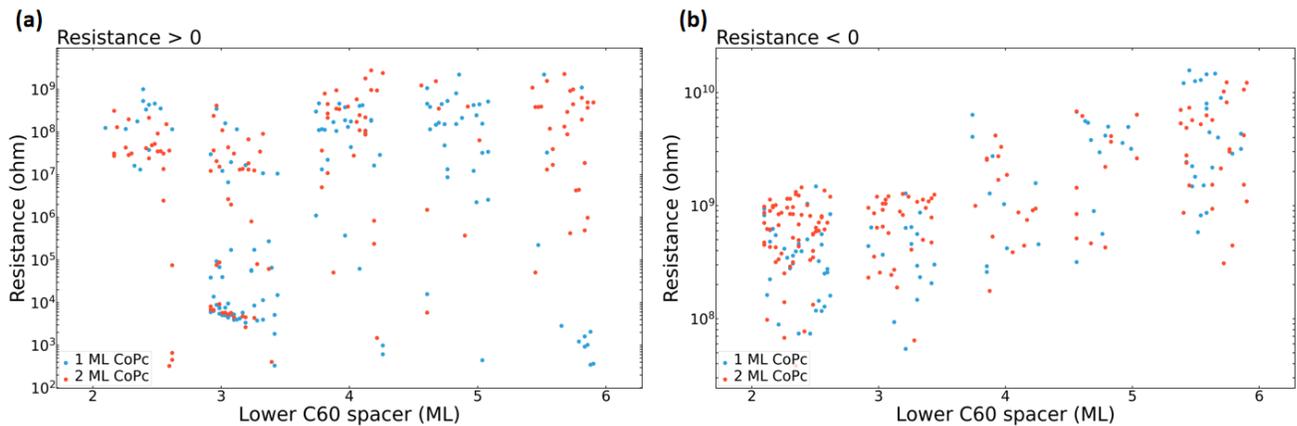

Figure 51: Junction prober data on TZ020425: Si/SiOx//Cr(5nm)/Fe(45nm)/$C_{60}$(2-6ML)/CoPc(1 or 2ML)/$C_{60}$(2-5ML)/Fe(15nm)/Cr(100nm). (a) Positive and (b) negative resistances (in Ohm) are shown. A Keithley 2636 sourcemeter with shielded cables up to the prober tips was used at ambient temperature and V=+5mV. Out of the 1414 devices, 887 (62.7%) were open-circuit, while 24 (1.7%) were short-circuited. The 503 working junctions correspond to a success rate of 35.6%. With measured currents at least one order of magnitude beyond experimental offsets, the negative resistance is interpreted as a signature of an active spintronic device[25].

The prober data of Figure 51 illustrates the nanojunction process's present capabilities. This data was acquired on TZ020425: Si/SiOx//Cr(5nm)/Fe(45nm)/$C_{60}$(2-6ML)/CoPc(1-2ML)/$C_{60}$(2-5ML)/Fe(15nm)/Cr(100nm). We display junctions that aren't open circuit (defined as I< $10^{-12}$ A or as exhibiting a charging behavior. Referring to the positive resistances measured, junctions with R<100 Ohm represent a short-circuit. A further batch of junctions have resistances in the $10^2$ < R (Ohm) < $10^6$ range. The majority of junctions lie in the range $10^7$ < R (Ohm) < $10^{10}$ range. In contrast, junctions with a negative resistance (i.e. electrically active) lie in the range $10^8$ < -R (Ohm) < $10^{10}$ range.

These data also demonstrate a small correlation between the molecular thickness of the molecular trilayer spacer and the resistance: 1) increasing the C60 thickness thanks to the two parallel wedges leads to over 1 order of magnitude increase in junction resistance. 2) There is a statistical increase in R when going from 1 to 2ML CoPc. This suggests that our wedge deposition capability is indeed modulating the effective junction thickness. To the best of our knowledge, whereas there exists limited data on the molecular thickness dependence of junction resistance [38], this represents the largest dataset to date. Given the large nominal variations in molecular layer thickness (from 2.1nm to 10.2nm nominally), one could reasonably expect a larger variation. This could be caused by imperfections in our wedge deposition capability. Indeed, to avoid the sample/mask clamps, the wedge shutter is positioned 3mm from the surface. This could cause shadowing issues. Also, the top electrode was deposited onto the sample while at room temperature, which can cause atomic metal interdiffusion. We are investigating solutions to both problems (see perspectives). As a final note, issues of organic residues that have not been removed from the nanopillars top might play a role in these statistics. Further experiments are needed to test this point.

# 11. Conclusions and Perspectives

In this technical paper, we have presented the iterations, trials and errors spanning ~10 years of engineering research, on a combined growth/technological process that can yield vertical molecular nanojunctions made from fully in-situ grown heterostructures. This advantageously allows the integration of water/air sensitive materials such as 3d ferromagnetic layers toward spintronic applications. Inspired by prior work on nanosphere lithography[48], we implemented several iterations of a process that uses silica nanobeads as a technological mask to design nanopillars between a junction's macroscopic electrodes. Although solvents are used as a microfluidic agent to position our beads in the most advanced iteration of the process, this solvent is deposited atop our heterostructure capping layer, so that this process otherwise does not expose the molecular layer to resists or solvents.

We described in detail the five technological steps, and additional precautions (e.g. attempts to precut the substrate prior to the process) to ensure the delivery of junctions on samples of the correct dimensions for magnetotransport experiments. Ultimately, we obtain a reasonable 'success' rate of junctions that are neither short-circuited nor open-circuit, with resistances at least one order of magnitude larger than that of the lower electrode access resistances. This does not, for now, lead to a Gaussian distribution of resistances around a mean value as can be observed for oxide junctions that are processed using industry-standard means (see e.g. Ref. [56]).

The process can be improved in the following ways:

1) improve the structural quality of the heterostructure deposition by reducing/eliminating the unwanted interdiffusion of metal atoms from the top ferromagnetic electrode into the molecular layer. We are presently testing a buffer-layer assisted growth[54,55] implementation on our multichamber cluster. In this technique, a buffer of Xe atoms is made to adsorb onto the molecular layer by cooling the sample to T<50K, so as to cushion the deposition of metallic atoms to form the ferromagnetic counterelectrode. The sample is then slowly warmed up to sublime the Xe atoms away, allowing the top electrode to 'land' onto the molecular layer and form what would become a high-quality top junction interface.
2) implement a true matrix printing array for which each microdroplet is positioned within the expected macro-area of a junction. This will help achieve better control over the junction's effective surface area and, if only 1 nanobead is present, make the junction's effective geometry conform to nominal expectations (i.e. not have several junctions in parallel). This goal should be within reach given our present technical capabilities.
3) reduce the junction's nominal size. Although 300nm beads were used for the dry dispersal process, 500nm beads were used for much of the auto-aligned process optimization. Given the process's thickness considerations (see Section 2.6), the 300nm node is definitely possible using the existing $SiO_2$ dielectric encapsulation. At the time of this writing, we have switched to a 300nm node. Looking ahead, it may be possible to use an atomic-layer deposition tool to encapsulate using high-density, stoichiometric $HfO_2$ or $Al_2O_3$ with a thickness as low as 10nm. If the thickness of other layers can be optimized, then in principle this suggests a possible ultimate junction diameter as low as ≈ 30nm.
4) As described in Section 10, the junction resistance as determined by 2-point probing appears to evolve with time. It is unclear what process is involved in this change, but a possible reason is the presence of unbound oxygen in the dielectric layer and its chemical interaction with the molecules along the pillar flanks. Switching to a nitride dielectric encapsulation layer could help address this issue.

As a mid-term perspective, our aim is to create an open-access materials science / technological platform with this process at its heart, in order to accelerate research into vertical molecular junctions. This should not only interest academics, who have by and large worked on shadow-mask macrojunctions, but also industrials who for the most part have shunned molecular nanojunction devices, and its potential uses in quantum technologies[25,26], because no such process existed. As a final note, this process does not require an e-beam writer, and thus has a lower technological footprint.

# 12. Acknowledgements

We thank B. Doudin for advice and microfluidic equipment. L.M.K. acknowledges Ph.D. fellowship received from 'Centre Franco-Indien pour la Promotion de la Recherche Avancée/Indo-French Center for the Promotion of Advanced Research (CEFIPRA/ IFCPAR)'. We acknowledge support from the 'CEFIPRA/ IFCPAR' via Project No. 5604-3, from the U. Strasbourg in 2016 (Espoirs award) and from the Contrat de Plan Etat Region of the French government in 2019. We acknowledge funding from the Agence Nationale de la Recherche (ANR-09-JCJC-0137, ANR-14-CE26-0009-01, ANR-21-CE50-0039), the Labex NIE 'Symmix' (ANR-11-LABX-0058 NIE), MICA Carnot 'Spinterface'; by "NanoTérahertz", a project cofunded by the ERDF 2014–2020 in Alsace (European Union Fund). This work of the Interdisciplinary Thematic Institute QMat, as part of the ITI 2021-2028 program of the University of Strasbourg, CNRS, and Inserm, was supported by IdEx Unistra (ANR 10 IDEX 0002) and by SFRI STRAT'US project (ANR 20 SFRI 0012) and EUR QMAT ANR-17-EURE-0024 under the framework of the French Investments for the Future Program. This work is supported by France 2030 government investment plan managed by the French National Research Agency under grant reference PEPR SPIN – SPINMAT ANR-22-EXSP-0007. We acknowledge the use of STNano facilities for device processing.